\documentclass[sigconf,arxiv,nonacm]{acmart}
\AtBeginDocument{%
  \providecommand\BibTeX{{%
    \normalfont B\kern-0.5em{\scshape i\kern-0.25em b}\kern-0.8em\TeX}}}

\setcopyright{acmcopyright}
\copyrightyear{2021}
\acmYear{2021}
\acmDOI{10.1145/1122445.1122456}

\acmConference[Montreal '21]{In Proceedings of
the 44th International ACM SIGIR Conference on Research and Development in Information Retrieval (SIGIR '21)}{July 11-15, 2021}{Virtual Event, Montreal, Canada}
\acmBooktitle{In Proceedings of
the 44th International ACM SIGIR Conference on Research and Development in Information Retrieval (SIGIR '21), July 11-15, 2021, Virtual Event, Montreal, Canada}
\acmPrice{15.00}
\acmISBN{978-1-4503-XXXX-X/18/06}

\usepackage{todonotes}
\usepackage{subfigure}

\acmSubmissionID{804}

\begin{document}

%%
%% The "title" command has an optional parameter,
%% allowing the author to define a "short title" to be used in page headers.

%\title{HeteGCN For Recommendation}
%\title{Heterogeneous Graph Convolutional Neural Networks \\ for Inductive Recommendation}
%\title{A Simple Neighborhood Aggregation Method Using Only User Embedding for Inductive Recommendation}
%\title{A Simple Neighborhood Aggregation Method for Inductive Recommendation}
\title{User Embedding based Neighborhood Aggregation Method \\ for Inductive Recommendation}
%\title{Embedding Items via Latent User Features: A Neighborhood Aggregation Method for Inductive Recommendation}
%\title{A Simple Neighborhood Aggregation Method for Inductive Recommendation by Embedding Items via Latent User Features}

%%
%% The "author" command and its associated commands are used to define
%% the authors and their affiliations.
%% Of note is the shared affiliation of the first two authors, and the
%% "authornote" and "authornotemark" commands
%% used to denote shared contribution to the research.
\author{Rahul Ragesh, Sundararajan Sellamanickam, Vijay Lingam*, Arun Iyer*, Ramakrishna Bairi*}\thanks{* Equal Contribution}
\email{t-rarage,ssrajan,t-vili,ariy,ram.bairi@microsoft.com}
\affiliation{%
  \institution{Microsoft Research India}
  \city{Bengaluru, India}
}

%%
%% By default, the full list of authors will be used in the page
%% headers. Often, this list is too long, and will overlap
%% other information printed in the page headers. This command allows
%% the author to define a more concise list
%% of authors' names for this purpose.
\renewcommand{\shortauthors}{R. Ragesh, et al.}
\newcommand{\bfx}{\mathbf{X}}
\newcommand{\bftx}{\mathbf{TX}}
\newcommand{\bff}{\mathbf{F}}
\newcommand{\bfn}{\mathbf{N}}
\newcommand{\bfl}{\mathbf{L}}
\newcommand{\bfa}{\mathbf{A}}
\newcommand{\bfd}{\mathbf{D}}
\newcommand{\bfta}{\tilde{\mathbf{A}}}
\newcommand{\bfh}{\mathbf{H}}
\newcommand{\bfg}{\mathbf{G}}
\newcommand{\bfe}{\mathbf{E}}
\newcommand{\bfz}{\mathbf{Z}}
\newcommand{\bfse}{\mathbf{e}}
\newcommand{\bfw}{\mathbf{W}}
\newcommand{\bfu}{\mathbf{U}}
\newcommand{\bfsu}{\mathbf{u}}
\newcommand{\bfv}{\mathbf{V}}
\newcommand{\bfsv}{\mathbf{v}}
\newcommand{\bfs}{\mathbf{S}}
\newcommand{\bfts}{\tilde{\mathbf{S}}}
\newcommand{\bftu}{\tilde{\mathbf{U}}}
\newcommand{\bfte}{\tilde{\mathbf{E}}}
\newcommand{\bfi}{\mathbf{I}}
\newcommand{\bfr}{\mathbf{R}}
\newcommand{\tgcn}{\textsc{TextGCN}\:}
\newcommand{\gcn}{\textsc{GCN}\:}
\newcommand{\hgcn}{\textsc{HeteGCN}\:}
\newcommand{\pte}{\textsc{PTE}\:}
\newcommand{\lgcn}{\textsc{LightGCN}\:}
\newcommand{\lr}{\textsc{LR}\:}
\newcommand{\cfu}{\textsc{CF-GCN-U}\:}
\newcommand{\cfe}{\textsc{CF-GCN-E}\:}
\newcommand{\cflu}{\textsc{CF-LGCN-U}\:}
\newcommand{\cfle}{\textsc{CF-LGCN-E}\:}
\newcommand{\cf}{\textsc{CF-GCN}\:}
\newcommand{\cfl}{\textsc{CF-LGCN}\:}
\newcommand{\ue}{\textsc{UE-GCN}\:}
\newcommand{\eu}{\textsc{EU-GCN}\:}

\newcommand{\bow}{\textsc{BoW}\:}
\newcommand{\tfidf}{\textsc{TF-IDF}\:}

\newcommand{\beginsupplement}{%
        \setcounter{table}{0}
        \renewcommand{\thetable}{S\arabic{table}}%
        \setcounter{figure}{0}
        \renewcommand{\thefigure}{S\arabic{figure}}%
     }

%%
%% The abstract is a short summary of the work to be presented in the
%% article.
\begin{abstract}
  We consider the problem of learning latent features (\textit{aka} embedding) for users and items in a recommendation setting. Given only a user-item interaction graph, the goal is to recommend items for each user. Traditional approaches employ matrix factorization-based collaborative filtering methods. Recent methods using graph convolutional networks (e.g., LightGCN) achieve state-of-the-art performance. They learn both user and item embedding. One major drawback of most existing methods is that they are not inductive; they do not generalize for users and items unseen during training. Besides, existing network models are quite complex, difficult to train and scale. Motivated by \lgcn, we propose a graph convolutional network modeling approach for collaborative filtering (\cf). We solely learn user embedding and derive item embedding using light variant (\cflu) performing neighborhood aggregation, making it scalable due to reduced model complexity. \cflu models naturally possess the inductive capability for new items, and we propose a simple solution to generalize for new users. We show how the proposed models are related to \lgcn. As a by-product, we suggest a simple solution to make \lgcn inductive. We perform comprehensive experiments on several benchmark datasets and demonstrate the capabilities of the proposed approach. Experimental results show that similar or better generalization performance is achievable than the state of the art methods in both transductive and inductive settings.
\end{abstract}

%%
%% The code below is generated by the tool at http://dl.acm.org/ccs.cfm.
%% Please copy and paste the code instead of the example below.
%%

%%
%% Keywords. The author(s) should pick words that accurately describe
%% the work being presented. Separate the keywords with commas.

% A "teaser" image appears between the author and affiliation
% information and the body of the document, and typically spans the
% page.
% \begin{teaserfigure}
%   \includegraphics[width=\textwidth]{sampleteaser}
%   \caption{Seattle Mariners at Spring Training, 2010.}
%   \Description{Enjoying the baseball game from the third-base
%   seats. Ichiro Suzuki preparing to bat.}
%   \label{fig:teaser}
% \end{teaserfigure}
% 

%%
%% This command processes the author and affiliation and title
%% information and builds the first part of the formatted document.
\maketitle

\section{Introduction}
Recommender systems design continues to draw the attention of researchers, as new challenges are to be addressed, with 
more demanding requirements coming due to several factors: (1) users (e.g., \textit{personalized} recommendation with high quality), (2) problem scale (e.g., number of users and items, rates at which they grow), (3) information available (e.g., history of user-item interactions, the volume of data, lack of side-information) for learning models \cite{Yehuda_MF_recommender_systems, LGCN}, and (4) system resources with constraints (e.g., storage cost, and inference speed). Since recommender systems provide\textit{personalized recommendation} because of their design, the development of collaborative filtering (CF) methods that make use of each user's past item interactions to build a model has drawn immense attention for more than a decade. CF methods' advantage is that they do not require other domain knowledge. Several important contributions \cite{CF_RecSys, CF_Survey1, CF_Survey2} have been made along dimensions including new modeling approaches \cite{Yehuda_MF_journal, Yehuda_MF_recommender_systems, NeuralCollaborativeFiltering, NAIS, LGCN}, optimization \cite{ALS_PR, GRMF, BPR, NegSamp}, scalability \cite{scalablecf_sru, disco, pardisCF, PinSage}. 

CF methods require \textit{only} user-item interaction graph to build models. Neighborhood and latent factor modeling methods are two important classes of CF methods \cite{Yehuda_MF_recommender_systems}. Neighborhood-based methods essentially work by discovering/computing relations between entities of the same types (e.g., item-item, user-user) and use this information to recommend items for users~\cite{fism}. Our interest lies in learning \textit{latent features} (a.k.a. embedding) for users and items. This way, user embedding can be directly compared with item embedding to score each user-item pair. Matrix factorization methods, a popular class of latent factor modeling methods, include Singular Value Decomposition \cite{SVD_Golub}, SVD++ \cite{Yehuda_MF_recommender_systems}, Alternating Least Squares (ALS) \cite{ALS_PR}, Factorization Machines (FM) \cite{Factorization_machines}. With the advent of neural modeling and deep learning methods, building recommender models using deep neural networks (DNN) and graph neural networks (GNN) has been surging. Neural models are very-powerful, and deliver high quality recommendations \cite{NeuralCollaborativeFiltering, NMF, Neural_Factorization_Machine_Predictive_Sparse_Analytics, NAIS}. However, they come with their challenges: high model complexity, training and inference costs, ability to work in the limited transductive setting, etc. See \cite{DL_based_recommender_system, DL_on_KG_recommender_system} for more details.  

Our focus in this work is to develop CF models using graph neural networks \cite{GCN, NeuralGraphCollaborativeFiltering, LGCN,  HierGCN, BGE} for learning user and item embedding. Graph convolutional networks (GCNs) have been used successfully in applications such as node classification, link prediction, and recommendation. Of particular interest here is to learn GCN models for collaborative filtering, when there is no side-information (e.g., node features, knowledge graphs). Several GCN model-based CF solutions \cite{NeuralGraphCollaborativeFiltering, LGCN, HierGCN} have been proposed recently. However, all of them have some limitations (e.g., high training cost, inability to handle large scale data, only transductive). We have two important requirements: (1) scale well - in particular when the number of items is large and significantly higher than the number of users; this situation is encountered in many real-world applications, and (2) able to generalize well on unseen users and items during inference, i.e., we need the model to be \textit{inductive}. 

Our work is inspired by a very recently proposed GCN model, \lgcn \cite{LGCN}. This important study investigated different aspects of GCNs for CF and proposed a very attractive simple model that delivered state-of-the-art performance on several benchmark datasets. However, it lacks two important requirements mentioned above. For example, \lgcn learns embedding for \textit{both} users and items. For reasons stated earlier, this model does not scale well, as the number of model parameters is dependent on the number of items. Therefore, it is expensive to train; also, it adds to model storage cost. Furthermore, \lgcn is transductive, as latent features are not available for new users and items during inference. We address these two important issues in this work, thereby expanding the capabilities of \lgcn. 
%Our approach is based on two key observations: (1) \lgcn showed that neighborhood aggregation is the most important function necessary in GCN, and (2) graph embedding based inductive recommendation model requires the ability to infer embedding for unseen users/items. 

A key contribution is the proposal of a novel graph convolutional network modeling approach for filtering (\cf), and we investigate several variants within this important class \cf models. The most important one is \cfu models which learn \textit{only} user embedding, and its \lgcn variant, \cflu (i.e., using only the most important function, neighborhood aggregation). Since \cflu models learn only user embedding, they can perform inductive recommendations with new items. However, this still leaves open the question: how to infer embedding for new users?. We suggest a simple but effective solution that answers this question. As a by-product, we also suggest a simple method to make LightGCN inductive. The \cflu modeling approach has several advantages. (1) The \cflu model has inductive recommendation capability, as it generalizes for users and items unseen during training. (2) It scales well, as the model complexity is dependent \textit{only} on the number of users. Therefore, it is very useful in applications where the growth rate of the number of items is relatively high. (3) Keeping both inductive and scalability advantages intact, it can achieve comparable or better performance than more complex models. 

 We suggest a twin-\cflu architecture to increase the expressive power of \cflu models by learning two sets of user embedding. This helps to get improved performance in some applications, yet keeping the model complexity advantage (with dependency only on the number of users). Furthermore, we show how the \cflu model and its counterpart (\cfle model that learns only item embedding) are related to \lgcn. Our basic analysis reveals how training the \cflu and \cfle models have interpretation of \textit{learning neighborhood aggregation functions} that make use of \textit{user-user} and \textit{item-item} similarities like neighborhood-based CF methods. 

We conduct comprehensive experiments on several benchmark datasets in both transductive and inductive setting. First, we show that the proposed \cflu models achieve comparable or better performance compared to LightGCN and several baselines on benchmark datasets in the transductive setting. Importantly, this is achieved with reduced model complexity, highlighting that further simplification of \lgcn is possible, making it more scalable. Next, we conduct an experiment to demonstrate the inductive recommendation capability of \cflu models. Our experimental results show that the performance in the inductive setting is very close to that achievable in the transductive setting. 

In Section 2, we introduce notation and problem setting. Background and motivation for our solution are presented in Section 3. Our proposal of \cf architecture and its variants are detailed in Section 4. We present our experimental setting and results in Section 5. This is followed by discussion and suggestions for future work, and related work in Sections 5 and 6. We conclude with important highlights and observations in Section 7.

\section{Notation and Problem Setting}

\textbf{Notation.} We use $\bfr \in \{0,1\}^{m \times n}$ to denote user-item interaction graph, where $m$ and $n$ denote the number of users and items respectively. Let $\bfu \in {\mathcal R}^{m \times d}$ and $\bfe \in {\mathcal R}^{n \times d}$ denote user and item embedding matrices.  We assume that the embedding dimension (d) is same for users and items. We use superscript to indicate the layer output embedding (e.g., $\bfu^{(l)}$ and $\bfe^{(l)}$ for $l^{th}$ layer embedding output of a network) explicitly, wherever needed. Lower-case letter and subscript are used to refer embedding vectors (row vectors of embedding matrices); for example, $\bfsu^{(l)}_j$ and $\bfse^{(l)}_k$ would denote embedding vectors of $j^{th}$ user and $k^{th}$ item at the $l^{th}$ layer output. 

\textbf{Problem Setting.} We are given only the user to item interaction graph, $\bfr$. We assume that no side information (e.g., additional graphs encoding item-item or user-user relations, the user or item features) is available. The goal is to learn latent features for users ($\bfu$) and items ($\bfe$) such that user-specific relevant recommendation can be made by ranking items using scores computed from embedding vectors (e.g., the inner product of user and item embedding vectors, $\bfz = \bfu \bfe^T$).

Most collaborating filtering methodologies cannot make recommendations for users/items unseen during training because they are transductive. Therefore, their utility is limited. We are interested in designing a graph embedding based \textit{inductive recommendation} model where we require the model to have the ability to generalize for users/items unseen during training. We assume that some interactions are available for new users/items during inference. Thus, we require an inductive recommendation modeling solution that can infer embedding vectors for new users and items. Moreover, in many practical applications, the number of items, $n$, grows much higher than the number of users, $m$. Therefore, we aim at developing models with reduced model complexity, having complexity independent of the number of items.    

\section{Background and Motivation}
%  We briefly present some details of these techniques and then make some observations to motivate our work.   

The simplicity and success of the recently proposed \lgcn model \cite{LGCN} for collaborating filtering inspire our work. \lgcn learns latent features (a.k.a. embedding) for users and items using a lightweight graph convolution neural network. We present some background on graph convolutional networks followed by \lgcn. We then make some observations to motivate our work. 

\textbf{Graph Convolutional Networks.} A graph convolutional network \cite{GCN} is composed using graph convolutional layers with each layer performing three basic operations: neighborhood aggregation, feature transformation, and non-linear activation. In GCN, $l^{th}$ layer function is defined as: $$\bfh^{(l+1)} = \sigma(\bfa\bfh^{(l)}\bfw^{(l)}).$$ It takes the previous layer output ($\bfh^{(l)})$ as input, transforms using weight matrix ($\bfw^{(l)})$, performs aggregation using an adjacency matrix ($\bfa$), and produces output ($\bfh^{(l+1)}$) via a nonlinear activation function $\sigma(\cdot)$ (e.g., ReLU or Sigmoid). Kipf and Welling \cite{GCN} proposed to use normalized $\bfa$ defined as: $\bfta = \bfd^{-\frac{1}{2}}(\bfi + \bfa)\bfd^{-\frac{1}{2}}$ where $\bfd$ is the diagonal in-degree matrix of $(\bfi + \bfa)$ and $\bfi$ helps to include self node representation in the neighborhood aggregation.  
%Each layer takes an input embedding matrix ($\bfh^{(l)}$) and produces an output embedding matrix ($\bfh^{(l+1)}$) using an adjacency matrix ($\bfa$) and a weight matrix ($\bfw^{(l)}$) as:  \bfw^{(l)})$ where $\bfg$ encodes relations among nodes in the graph and $\sigma(\cdot)$ denotes some nonlinear activation function (e.g., ReLU or Sigmoid). 
In node classification tasks, $\bfh^{(0)}$ comprises of node features and the model weights $\{\bfw^{(l)}: l=0,\ldots,L\}$ are learned by optimizing an objective function (e.g., cross-entropy loss) using labeled data. 

\textbf{Recommendation Setting.} In the \textit{basic} model with GCN, we use  
 %\begin{equation}
 $\bfa_{CF} = \begin{bmatrix}
 {\bf 0} & \bfr \\
 \bfr^T & {\bf 0}
 \end{bmatrix}$ 
 as the graph 
 %\label{eqn:G}
%\end{equation}
and $\bfh^{(l)} = \begin{bmatrix} \bfu^{(l)} \\ \bfe^{(l)} \end{bmatrix}$. Note that user and item embedding vectors are available at each layer output, and propagated through multiple layers. Then, user and item embedding (starting with 1-hot representation at the input) are learned with the GCN model weights using a loss function suitable for collaborative filtering. Following \lgcn \cite{LGCN}, we use Bayesian personalized ranking (BPR) loss function defined as:
\begin{equation}
    {\mathcal L} = \sum_{i=1}^m \sum_{j \in \mathcal{N}_e(u_i)} \sum_{k \in NS_e(u_i)} g(z_{i,j},z_{i,k};\bfh) + \lambda ||\bfh||^2
    \label{eqn:trloss}
\end{equation}
where $\mathcal{N}_e(u_i)$ and $NS_e(u_i)$ denote item neighborhood (with interactions) and negative samples (items) of user $u_i$. The function $g(\cdot)$ measures degree of violation of positive pairwise score ($z_{i,j}$) higher than negative pairwise score ($z_{i,k}$) and we use $g(z_i,z_j) = -\ln \sigma(z_{i,j}-z_{i,k})$ in our experiments. $\lambda$ is weight regularization constant and $||\bfh||^2$ denote L2-norm of embedding and layer weight vectors. We give more details of model training in the experiment section. 

\textbf{\lgcn.} 
%\textit{Apart from several fundamental differences between node classification and recommendation problems (e.g., loss functions used for training, performance metrics, availability of labeled and unlabeled data), one important difference in our setting is that we do not have $\bff$ which is equivalent to having side information as user-user relational graph and this makes the problem lot more challenging, as we discuss shortly.}    
\lgcn draws upon the idea of propagating embedding from prior work on neural graph collaborative filtering (NGCF) \cite{NeuralGraphCollaborativeFiltering}. NGCF uses the basic GCN recommendation model explained above except that an additional term involving element-wise (Hadamard) product of user and item embedding is part of neighborhood aggregation. See \cite{NeuralGraphCollaborativeFiltering} for more details. He et.al. \cite{NeuralGraphCollaborativeFiltering} empirically showed that feature transformation and non-linear activation operations are unnecessary in NGCF and can even be harmful to the performance. They proposed a simpler GCN architecture (\lgcn) by keeping only neighborhood aggregation function with the result:  $\bfh^{(l+1)} = \bfta \bfh^{(l)}, l = 1,\ldots,L$. \lgcn fuses these multi-layer outputs as:
\begin{equation}
\bfh = \sum_{l=0}^L \alpha_l \bfh^{(l)} = \sum_{l=0}^L \alpha_l \bfta^l \bfh^{(0)}
\label{eqn:LGCN}
\end{equation}
where $\{\alpha_l: l = 0, \ldots, L\}$ are fusion hyperparameters. It is worth noting that \lgcn does not require $\bfi$ while using $\bfta$, as defined by Kipf and Welling \cite{GCN}. Also, note that since $\bfr$ is a bipartite graph, user and item specific in-degree matrices ($\bfd_u$ and $\bfd_e$) are used in pre/post multiplication of $\bfa$. See \lgcn \cite{LGCN} for more details. In empirical studies, they found that tuning fusion hyperparameters is not necessary and setting equal weights (i.e., simple average) works well. From (\ref{eqn:LGCN}), we see that the only learnable parameters are $\bfh^{(0)} = \begin{bmatrix} \bfu^{(0)} \\ \bfe^{(0)} \end{bmatrix}$. Furthermore, higher order neighborhood aggregation happens through the effective propagation matrix, $\sum_{l=0}^L \alpha_l \bfa^l$. Experimental results \cite{LGCN} showed state-of-the-art performance achievable by \lgcn on several benchmark datasets. 

\textbf{Motivation.} We observe that \lgcn is quite attractive due to its simplicity and ability to give a state-of-the-art performance. However, it still has two drawbacks. Firstly, it is transductive. Therefore, new items, when they arrive, cannot be recommended for existing users. Similarly, recommending existing items to new users is not possible. Secondly, the model complexity of \lgcn is $O((m+n)d)$. Thus, it poses a scalability issue, for example, when the number of items is extremely huge (i.e., $n \gg m$). In many real-world applications, the rate at which the number of items grows is significantly higher compared to users. Thus, our interest lies in answering two key questions: 
\begin{itemize}
    \item \textbf{RQ1:} \textit{How do we make GCN models for collaborative filtering (\cf) scalable, make it independent of the number of items, yet delivering state-of-the-art performance?} 
    \item \textbf{RQ2:} \textit{How do we make \cf models inductive?}
    %\item \textbf{RQ3:} \textit{Is it possible to make LightGCN inductive? If yes, how?}
\end{itemize}
 We answer these questions in the next section. As a by-product, we also show how \lgcn can be made inductive. 
 %Our approach is motivated by a recently proposed \hgcn modeling approach \cite{HeteGCN} which addresses similar issues in small-labeled document classification setting. However, it is important to note that node features are available and used in text/node classification tasks, and this is not the case in the collaborative filtering setting.  

%by proposing a new architecture, CF-GCN-U and its light version, CF-LGCN-U where the letters \textbf{CF}, \textbf{U} and \textbf{CF} refer to using bipartite graph ($\bfr$) in the GCN layer, learning only user embedding vectors ($\bfu^{(0)}$) and a light version similar to the original LightGCN model, respectively.       

%\begin{enumerate}
%    \item Introduce Notation 
%    \item State the problem 
%    \item Motivate with Model Simplification (connecting with LightGCN), scalable and inductive Learning 
%\end{enumerate}

\section{Proposed Solution}
%The proposed architecture is motivated by the recent work on heterogeneous graph neural networks (HeteGCN) of Rahul Ragesh et. al. [] developed for text classification problems. We briefly explain HeteGCN first before proposing our architecture. 
%We drew upon an analogy between the TextGCN and CF-GCN network architectures in Section 3, and highlighted how HeteGCN addresses scalability and transductive issues of TextGCN. Further, we observed that similar issues arising in the recommendation setting as well. Then, the natural question that arises is: \textit{what is the HeteGCN equivalent architecture for collaborative filtering?} We answer this question in this question. 

%We present our central idea of learning \textit{only} user embedding and deriving item embedding using a new architecture, \cfu and highlight its connection with the \lgcn solution. We propose a new architecture, \cfu for the collaborative filtering problem. %However, there are several important differences and non-trivial aspects to handle compared to the classification setting. 
%We discuss several variants of the proposed architecture and how simplifying using \lgcn idea results in lighter models, \cflu. After establishing the simplicity in terms reduced model complexity for the proposed architecture, we show how to use the proposed architecture for the inductive recommendation setting. Finally, we also suggest a simple way of making \lgcn inductive.

We propose a novel graph convolutional network modeling approach for \textit{inductive} collaborative filtering (\cf). We present the idea of learning \cf models with \textit{only} user embedding (termed \cfu and its \textit{light} version, \cflu) and infer item embedding. We discuss several \cf model variants that can be useful for different purposes (e.g., increasing the expressive power of models, scenarios where learning item embedding is beneficial). With basic analysis, we show how the proposed models are related to \lgcn. We also show that training these models can be interpreted as learning neighborhood aggregation function using user-user and item-item similarity function like neighborhood-based CF methods. 
%After establishing the simplicity in terms reduced model complexity for the proposed architecture, 
We show how to use the proposed architecture for the inductive recommendation setting. Finally, we also suggest a simple way of making \lgcn inductive.

\begin{figure*}%
        \includegraphics[width=0.65\textwidth]{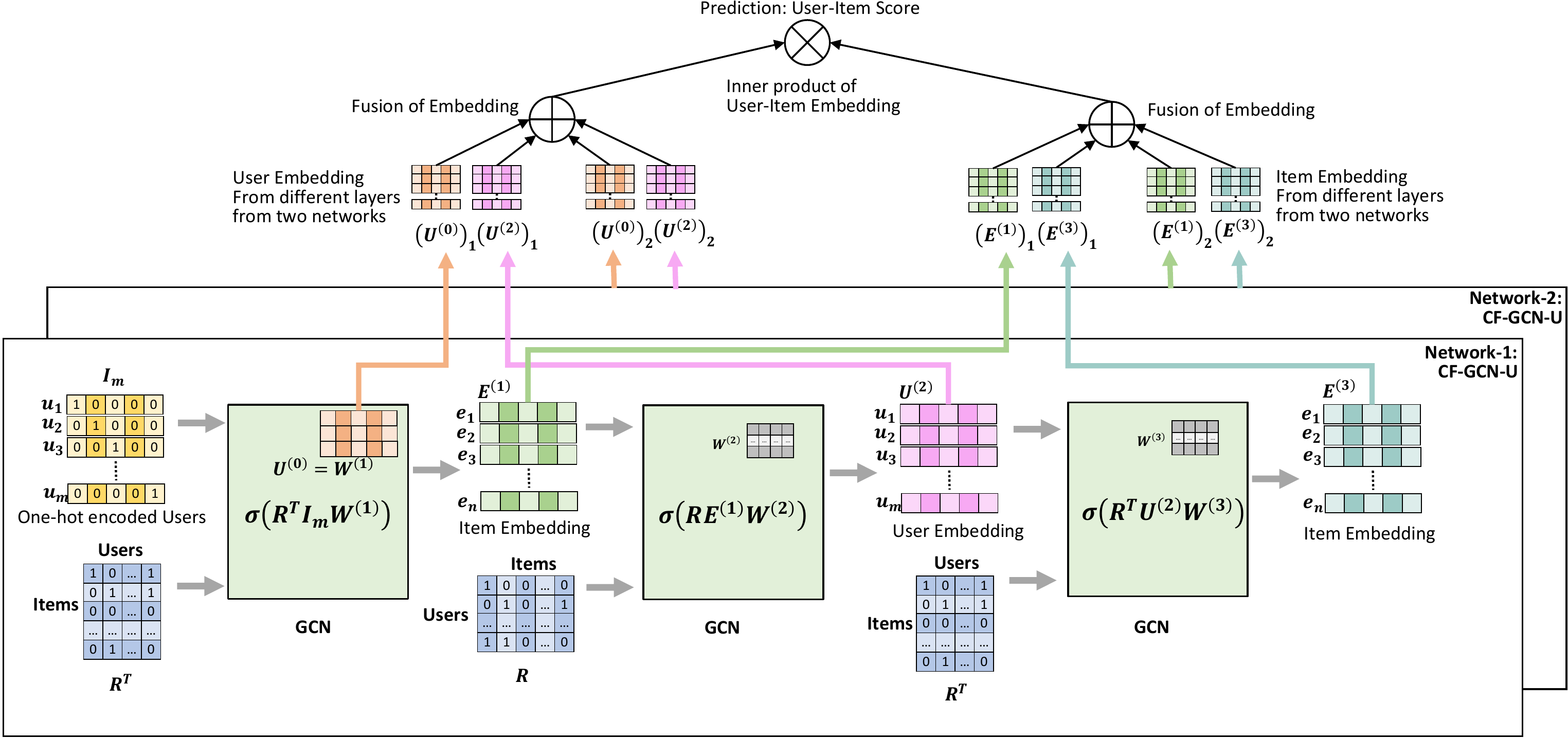}
        \caption{\label{fig:Architecture} \small{Graph Convolutional Network for Collaborating Filtering Architecture (\cfu) with Twin Networks. Each network is \cfu, using \gcn layers with graphs $\bfr$ and $\bfr^T$. We learn user embedding and derive item embedding. Fuse user and item embedding from different layers/networks. Removing layer weight matrices (except in the first layer) and non-linear activation function $\sigma(\cdot)$ result in the \cflu network.}}
\vspace{-5mm}
\end{figure*}

%\subsection{Background}
  
 \subsection{Collaborating Filtering Networks}
 The basic idea that we take~\cite{HeteGCN} is to use heterogeneous graph convolutional layers to compose individual networks and combine embedding outputs from multiple layers and networks. In our setting, since we have only two graphs, $\bfr$, and $\bfr^T$, there can only be two-layer types. We define \ue layer as the GCN layer that takes item embedding ($\bfe$) as input and uses $\bfr$ to produce user embedding ($\bfu$) output. A similar definition holds for \eu layer. 
 \vspace{-3mm}
 \subsubsection{\cf Networks} 
 A \cf network is composed of a cascaded sequence of \ue and \eu layers. There are only two network types for two reasons: (1) we have only two embedding types (user and item) and start with only one embedding input (i.e., user or item), and (2) we have only two graphs ($\bfr$ and $\bfr^T$) and layer compatibility (i.e., output and input of two consecutive layers match) is to be ensured. We name these networks, \cfu and \cfe, as they take user and item embedding input respectively. 
 
  We illustrate with an example. \cf($\bfr^T$-$\bfr$-$\bfr^T$) is a three-layer \cfu network, starting with user embedding input ($\bfu^{(0)}$). See Figure~\ref{fig:Architecture}. It produces item embedding outputs, $\bfe^{(1)} = \sigma(\bfr^T \bfi_m \bfw^{(1)})$  and $\bfe^{(3)} = \sigma(\bfr^T \bfu^{(2)} \bfw^{(3)})$, as the first and third layer outputs respectively. Note that $\bfu^{(0)} = \bfw^{(1)}$ (with 1-hot encoding $\bfi_m$ for users). Further, the network produces user embedding, $\bfu^{(2)} = \sigma(\bfr \bfe^{(1)} \bfw^{(2)})$, as the second layer output. We fuse $\bfu^{(0)}$ and $\bfu^{(2)}$ using AGGREGATION (e.g., mean or weighted sum) or CONCAT functions. Likewise, we obtain a fused item embedding. Finally, we compute each user-item pairwise score, as the inner product of user and item embedding vectors. In this network, the learnable parameters are $\bfu^{(0)}$ and GCN layer weights ($\bfw^{(2)}$ and $\bfw^{(3)})$. Thus, we learn \textit{only} user embedding, and derive item embedding using this architecture. Likewise, \cf($\bfr$-$\bfr^T$-$\bfr$) is a three layer \cfe network with learnable $\bfe^{(0)}$ (\textit{only} item embedding) and layer weights.
 
 %\textbf{\cfl Networks.} 
 \vspace{-1mm}
 \subsubsection{\cfl Networks}
 As explained earlier, \lgcn uses only neighborhood aggregation and abandons feature and nonlinear transformations. Using this idea, we simplify $\cf$ networks. Let us continue with the three-layer \cfu network example. On dropping feature and nonlinear transformation, we get the \cflu network with outputs as:
 \begin{eqnarray}
     \bfe^{(1)}_u = \bfr^T \bfu^{(0)},  \bfe^{(3)}_u = \bfr^T \bfs_u \bfu^{(0)} \label{eqn:eemb_lgcnu} \\
     \bfu^{(2)}_u = \bfs_u \bfu^{(0)} \label{eqn:uemb_lgcu}
 \end{eqnarray}
  where $\bfs_u = \bfr \bfr^T$ measures user-user similarity. Note that we use the subscript $u$ to denote \textit{only user} embedding model. Using weighted mean with weights $(\alpha_0, \alpha_1)$ to fuse, we get:
  \begin{eqnarray}
      \bfe_u = \bfr^T (\alpha_0 \bfi_m + \alpha_1 \bfs_u) \bfu^{(0)} \label{eqn:eemb_lgcnu2} \\
      \bfu_u = (\alpha_0 \bfi_m + \alpha_1 \bfs_u) \bfu^{(0)}. \label{eqn:uemb_lgcnu2}
  \end{eqnarray}
  Note that we have used same set of weights (tunable hyperparameters) in (\ref{eqn:eemb_lgcnu2}) and (\ref{eqn:uemb_lgcnu2}) for simplicity, and we can also use layer-wise weights (e.g., ($\alpha_1$ and $\alpha_3$ in \ref{eqn:eemb_lgcnu2}) and ($\alpha_0$ and $\alpha_2$ in \ref{eqn:uemb_lgcnu2})). Thus, \cflu learns \textit{only user} embedding, and uses second order user-user information ($\bfs_u$) to infer final user embedding ($\bfu_u$) that captures user-user relation as well. Further, item embedding is inferred by aggregating over its neighbor user embedding (see (\ref{eqn:eemb_lgcnu2})). It is also useful to note that since $\bfs_u$ is obtained using \textit{only} $\bfr$, we have: $\bfr^T \bfs_u = \bfs_e \bfr^T$ where $\bfs_e = \bfr^T \bfr$. Thus, we can rewrite $\bfe^{(3)}$ in (\ref{eqn:eemb_lgcnu}) and $\bfe_u$ in (\ref{eqn:eemb_lgcnu2}) as: 
   \begin{equation}
       \bfe^{(3)}_u = \bfs_e \bfr^T \bfu^{(0)}, \bfe_u = (\alpha_0 \bfi_n + \alpha_1 \bfs_e) \bfr^T \bfu^{(0)}.
       \label{eqn:eemb_lgcnu3}
   \end{equation}
 With the definition, ${\hat \bfe^{(0)}_u} = \bfr^T \bfu^{(0)}$, we can interpret ${\hat \bfe^{(0)}_u}$ as an effective or proxy item embedding derived from the learned user embedding, with the result: $\bfe_u = \bfts_e {\hat \bfe^{(0)}_u}$ where $\bfts_e = (\alpha_0 \bfi_n + \alpha_1 \bfs_e)$. 
 
 It is useful to understand the number of users and item embedding sets we get from a multi-layer network. In \textit{one} layer case, we have one set of item and user embedding, i.e., $\bfu^{(0)}$ and $\bfe^{(1)}$. With \textit{two} layers, we get two sets of user embedding and one set of item embedding. With \textit{three} layers, we have two sets each, as seen from (\ref{eqn:eemb_lgcnu2}) and (\ref{eqn:uemb_lgcnu2}). To generalize, we get one set of item embedding \textit{lesser} with \textit{even} number of layers. It does not pose any issues in mean fusion. However, when we have even number of layers and use inner product score, we need to drop one set of embedding (either user or item) to ensure dimensions match with \textsc{concat} fusion.     
 
 Generalizing (\ref{eqn:uemb_lgcnu2}) and (\ref{eqn:eemb_lgcnu3}) with more layers, fusing with weighted mean and using inner product score ($\bfz_u = \bfu_u \bfe^T_u)$, we get the multi-layer \cflu network (with $2L+1$ layers) outputs as:  
  \begin{eqnarray}
      %\bfe_u = \bfr^T \bfts_u \bfu^{(0)} = \bfts_e {\hat \bfe^{(0)}} \label{eqn:eemb_lgcnu3} \\
      %\bfe_u = \bfts_e {\hat \bfe^{(0)}}, \bfu_u = \bfts_u \bfu^{(0)} \label{eqn:lgcnu4} \\
      \bfe_u = \bfr^T \bfts_u \bfu^{(0)}, \bfu_u = \bfts_u \bfu^{(0)} \label{eqn:lgcnu4} \\
      %\bfu_u = \bfts_u \bfu^{(0)} \label{eqn:uemb_lgcnu3} \\
      %\bfz_u = \bfts_u \bfu^{(0)} ({\hat \bfe^{(0)}_u})^T \bfts^T_e \label{eqn:zuscore}
       \bfz_u = \bfts_u \bfu^{(0)} (\bfu^{(0)})^T \bfts^T_u \bfr \label{eqn:zuscore}
  \end{eqnarray}
  where $\bfts_u = \sum_{l=0}^L \alpha_l \bfs^l_u$, $\bfs^{0}_u = \bfi_m$. Thus, zero and higher powers of user-user and item-item similarities are used with increase in the number of layers. Applying the same steps to the \cfle network, we get:
  %$\bfts_e = \sum_{l=0}^L \alpha_l \bfs^l_e$, $\bfs^{0}_e = \bfi_n$; here, we assume $L$ to be odd. Thus, zero and higher powers of user-user and item-item similarities are used with increase in the number of layers. Applying the same idea to the \cfle network, we get:  
  \begin{eqnarray}
      %\bfe_e = \bfts_e \bfe^{(0)}, \bfu_e = \bfr \bfts_e \bfe^{(0)} = \bfts_u {\hat \bfu^{(0)}} \label{eqn:eemb_lgcne2} \\
      %\bfe_e = \bfts_e \bfe^{(0)}, \bfu_e = \bfts_u {\hat \bfu^{(0)}} \label{eqn:eemb_lgcne2} \\
      \bfe_e = \bfts_e \bfe^{(0)}, \bfu_e = \bfr \bfts_e \bfe^{(0)} \label{eqn:eemb_lgcne2} \\
      %\bfu_e = \bfr \bfts_e \bfe^{(0)} = \bfts_u {\hat \bfu^{(0)}} \label{eqn:uemb_lgcne2} \\
      \bfz_e = \bfr \bfts_e \bfe^{(0)} (\bfe^{(0)})^T \bfts^T_e \label{eqn:zescore}
  \end{eqnarray}
  where $\bfts_e = \sum_{l=0}^L \alpha_l \bfs^l_e$, $\bfs^{0}_e = \bfi_n$. %${\hat \bfu^{(0)}} = \bfr \bfe^{(0)}$. 
  This network uses \textit{only item} embedding as denoted by the subscript $e$ and, higher-order similarity information to learn/infer item/user embedding.   
  \vspace{-1mm}
  \subsubsection{Interpretation of \cflu and \cfle learning} 
  We observe that (\ref{eqn:zuscore}) can be rewritten as: $\bfz_u = \bfw_u \bfr$, where $\bfw_u = \bfts_u \bfu^{(0)} (\bfu^{(0)})^T \bfts^T_u$. 
  %and this relation is obtained from noting that $\bfr^T \bfts_u = \bfts_e \bfr^T$, $\bfts_e {\hat \bfe^{(0)}_u} = \bfr^T \bfts_u \bfu^{(0)}$ and  ${\hat \bfe^{(0)}} = \bfr^T \bfu^{(0)}$. 
  Thus, we can interpret training as learning user-centric aggregation function (expressed through learnable user embedding, with a special structure), $\bfw_u$,  that takes into account the user-user similarity. Likewise, the scoring function with \cfle (\ref{eqn:zescore}) can be rewritten as: $\bfz_e = \bfr \bfw_e$, where $\bfw_e = \bfts_e \bfe^{(0)} (\bfe^{(0)})^T \bfts^T_e$. Thus, the score is computed using an item centric transformation (post multiplication) or aggregation function that uses item-item similarity. Thus, \cflu and \cfle offer learning user and item oriented neighborhood aggregation functions for collaborative filtering.  
  
  One natural question is: \textit{Which network is better - \cflu or \cfle?}. We conduct a comparative experimental study later. Note that the number of model parameters is more for \cfle when $n > m$. These additional degrees of freedom may help in some applications to get improved performance. However, we emphasize that our interest primarily lies in \cflu networks only due to reasons explained earlier. We include \cfle discussion to show it as a possible variant within the class of \cf models. Furthermore, it helps to connect \cf modeling approach with \lgcn, as explained below. 
 
 \subsection{Model Complexity}
We observe that the standard GCN model, including \lgcn, uses $\bfa$ as the adjacency matrix in every GCN layer. Thus, it produces both user and item embedding as each layer output, having $O((m+n)d)$ model complexity. On the other hand, the model complexities of \cfu and \cfe networks are $O(md)$ and $O(nd)$ only. Thus, these networks help to reduce model complexity. It is possible to train \cfu and \cfe networks jointly, and fuse user/item embedding outputs from \textit{both} networks. However, this solution has $O((m+n)d)$ complexity. \lgcn is one such solution, as we show next.  %As explained earlier, since the number of items grow higher rate than the number of users we conduct comprehensive experiments with \cfu networks.

\subsection{Relation with LightGCN}
To understand the relation with LightGCN, we begin with the expression (\ref{eqn:LGCN}) for the embedding vectors, and unfold up to three layers as follows:   
\begin{equation}
 \bfa^2_{CF} = \begin{bmatrix}
 \bfs_u & \bf{0} \\
 \bf{0} & \bfs_e
 \end{bmatrix}, 
 \bfa^3_{CF} = \begin{bmatrix}
 \bf{0} & \bfs_u\bfr \\
 \bfs_e \bfr^T & \bf{0} 
 \end{bmatrix}
 \label{eqn:ALGCN2}
\end{equation}
where $\bfs_u = \bfr\bfr^T$ and $\bfs_e = \bfr^T\bfr$, and they measure user-user and item-item similarities.   
%\begin{equation}
% \bfa^3_{\lgcn} = \begin{bmatrix}
% \bf{0} & \bfs_u\bfx \\
% \bfs_e \bfx^T & \bf{0} 
% \end{bmatrix}
% \label{eqn:ALGCN3}
%\end{equation}
Note that while $\bfa$ is off-block-diagonal, $\bfa^2$ is block-diagonal. Thus, there is a special structure to odd and even powers of $\bfa$. Using (\ref{eqn:LGCN}) and (\ref{eqn:ALGCN2}), we get: 
\begin{eqnarray}
    \bfu = (\alpha_0 \bfi_m + \alpha_2 \bfs_u) \bfu^{(0)} + \bfr (\alpha_1 \bfi_n  + \alpha_3 \bfs_e) \bfe^{(0)} \label{eqn:uemb_lgcn3}\\
    \bfe = \bfr^T (\alpha_1 \bfi_m + \alpha_3 \bfs_u) \bfu^{(0)} + (\alpha_0 \bfi_n + \alpha_2 \bfs_e) \bfe^{(0)} \label{eqn:eemb_lgcn3}
\end{eqnarray}
%and can be rewritten in matrix form as:
%\begin{eqnarray}
%\bfh^{(3)} = \begin{bmatrix} \bfu^{(3)} \\ \bfe^{(3)} \end{bmatrix} = \begin{bmatrix} (\alpha_0 \bfi + \alpha_2 \bfs_u) & \bfx (\alpha_1 \bfi + \alpha_3 \bfs_e) &  \\ \bfx^T (\alpha_1 \bfi + \alpha_3 \bfs_u) & (\alpha_0 \bfi + \alpha_2 \bfs_e) \end{bmatrix} \begin{bmatrix} \bfu^{(0)} \\ \bfe^{(0)} \end{bmatrix}
%\end{eqnarray}
On comparing (\ref{eqn:uemb_lgcn3}), (\ref{eqn:eemb_lgcn3}) with three layer embedding outputs of \cflu (see (\ref{eqn:eemb_lgcnu2}), (\ref{eqn:uemb_lgcnu2})) and similar expressions for \cfle networks (not shown), we see that \lgcn combines these two network embedding outputs with suitable matching weights. Thus, it makes use of \textit{both} user-user and item-item second order information in computing user and item embedding. Further, the inner product score computed by \lgcn using (\ref{eqn:uemb_lgcn3}) and (\ref{eqn:eemb_lgcn3}) is essentially sum of the \cflu and \cfle scores (i.e., (\ref{eqn:zuscore}) and (\ref{eqn:zescore})); in addition, it has \textit{cross-term} inner product scores of user embedding of \cflu (i.e., the first term in (\ref{eqn:uemb_lgcn3})) with item embedding \cfle (i.e., the second term in (\ref{eqn:eemb_lgcn3})) and vice-versa.  

\subsection{Twin \cflu Networks} As noted earlier, \lgcn uses more parameters and has model complexity $O((m+n)d)$; therefore, more powerful. %Also, it is not clear whether the cross-terms help. 
The question is: \textit{can \cflu match \lgcn performance with such reduced model complexity?}. One simple way to increase the power of \cflu is to use two (twin) \cflu networks, each learning different sets of user embedding (i.e., $\bfu^{(0)}_1$ and $\bfu^{(0)}_2$). In this case, the effective item embedding of the second network is: $(\bfe^{(0)}_u)_2 = \bfr^T \bfu^{(0)}_2$ and has more degrees of freedom compared to the single \cflu network where the effective item embedding is dependent only on $\bfu^{(0)}$. Note that we can have different number of layers in each network, and the user/item embedding from both networks are fused to get the final user/item embedding. The advantage of the twin \cflu network modeling approach is that the model complexity is still independent of the number of items and significant reduction is achieved when $n \gg m$. 

The other question is: \textit{do we need both networks to get state-of-the-art performance?.} In the experiment section, we evaluate different \cf networks on several benchmark datasets. We show that \cflu network gives a comparable or better performance than \lgcn through empirical studies. Furthermore, the model complexity is reduced and incurs lesser storage costs when $m < n$.  

%We observe that there are two key matrices, $\bfts_u = \sum_{l=0}^L \bfs^{l}_u$ and $\bfts_e = \sum_{l=0}^L \bfs^{l}_e$, and transformed embedding matrices, $\bftu = \bfts_u \bfu^{(0)}$ and $\bfte = \bfts_e \bfe^{(0)}$ where transformation projects using user-user and item-item similarities. (XXX Some check is needed for indices.) This results in: $\bfu = \alpha (\bftu + \bfx \bfte)$ and $\bfe = \alpha (\bfx^T \bftu + \bfte)$.   
%Our first observation is that $\bfa$ has only off-diagonal matrices. 
%and split into user and item embedding matrices, that is: $\bfu = \sum_{l=0}^L \alpha_l \bfa^l \bfu^{(0)}$ and $\bfe = \sum_{l=0}^L \alpha_l \bfa^l \bfe^{(0)}$. Let $\bfp = \sum_{l=0}^L \bfa^l$. 
%In LightGCN, score for each user-item pair is computed as inner product, $\bfz = \bfu \bfe^T$. 
\vspace{-1mm}
\subsection{Inductive Recommendation} 
\lgcn is a transductive method because we learn \textit{both} user/item embedding and it is not apparent how to infer embedding for new users and items. Graph embedding based inductive recommendation model requires the ability to infer new user and item embedding. Using \cfl networks which learn only user embedding meets this requirement as discussed below. We present our solution starting with the easier problem of fixed user set with training data available to learn user embedding and we need to generalize only for new items. Then, we expand the scope to include new users and suggest a simple solution for \cflu networks. Also, as a by-product, we show how \lgcn can be modified for making inductive inference. 
\vspace{-1mm}
\subsubsection{Inductive \cflu: Only New Items} Let us consider the three-layer \cflu example again (see (\ref{eqn:eemb_lgcnu}) - (\ref{eqn:eemb_lgcnu3})). Since we learn only user embedding and infer item embedding, \cflu can infer embedding for new items. Let $\bfr_I$ denote \textit{new} user-item interaction graph such that new items (i.e., new columns) are also appended. We assume that at least a few interactions are available for each new item in $\bfr_I$. In practice, this is possible because one can record interactions on new items shown to randomly picked or targeted existing users. Recall that the three-layer \cflu network uses $\bfr^T-\bfr-\bfr^T$ in tandem. We compute new item embedding by substituting $\bfr_I$ for $\bfr$ in each layer. Note that we can choose to use $\bfr$ or $\bfr_I$ in the second layer (i.e., to compute fresh user embedding or not). We conduct an ablation study to evaluate the efficacy of using fresh user embedding and present our results shortly. 
\vspace{-1mm}
\subsubsection{Inductive \cflu: New Users} Since we learn user embedding in \cflu, we face the same question of \textit{how to infer embedding for new users?.}  When we look at (\ref{eqn:eemb_lgcnu2}) and (\ref{eqn:uemb_lgcnu2}) closely, the main difficulty arises from the term $\alpha_0 \bfi_m \bfu^{(0)}$, as we cannot get embedding for new users without retraining. We propose a naive solution to address this problem. Suppose we set $\alpha_0 = 0$, that is, we do not use $\bfu^{(0)}$ in the fusion step. We can then still compute fresh user embedding by substituting new $\bfr_U$ (i.e., $\bfr$ with added rows) in the second layer. However, the number of user embedding sets available for fusion is lesser by one set. Since the inner product of fused user and item embedding requires that the dimensions match, we also drop the first layer item embedding output and add one more layer, if needed, to ensure a valid \textsc{CONCAT} fusion. This simple solution provides the ability to infer new user embedding, and we demonstrate its usefulness through empirical studies. 
\vspace{-1mm}
\subsubsection{Inductive \cflu: New Users and Items} We handle both new users/items' by not using embedding outputs of the first layer (i.e., by setting $\alpha_0 = 0$ for the user and item embedding, as needed). Note that we update \textit{both} user and item embedding. This update requires using $\bfr_I$ (i.e., ignoring new users) in the first layer and fully updated, $\bfr$ (i.e., having both new users and items) in other layers.. It is important to keep in mind that we require at least a few labeled entries for new users/items during inductive inference. Finally, we note that the idea of not using the first layer user embedding is useful for \lgcn. Our experimental results show that \textit{inductive} \lgcn also works. 
\vspace{-1mm}
\section{Experiments}
We first discuss experimental setup, including a brief description of the datasets and introduces the baselines used; this sub-section also covers the training details, metrics and hyperparameter optimization. We evaluate our proposed approach in both Transductive and Inductive settings, present and discuss our results.  

\vspace{-1mm}
\subsection{Experimental Setup}
\subsubsection{Datasets}
We used the three datasets provided in the LightGCN repository \footnote{https://github.com/kuandeng/LightGCN} (a) \textbf{Gowalla}~\cite{gowalla} contains the user location check-in data (b) \textbf{Yelp2018}~\cite{NeuralGraphCollaborativeFiltering} contains local business (treated as item) recommendation to users (c) \textbf{Amazon-book}~\cite{Amazon} contains book recommendation to users. We include one additional dataset \textbf{Douban-Movie}~\cite{douban} to our evaluation; this dataset consists of user-movie interactions. The statistics of datasets are detailed in the table \ref{tab:datasetStats}.
\vspace{-1mm}
\subsubsection{Transductive Dataset Preparation}
As the repository only contained train and test splits, we sampled 10\% of items randomly for each user from the train split to construct the validation split. As there is a difference in the splits used in LightGCN, we retrain all the baselines on these new splits and report the metrics. 
\vspace{-1mm}
\subsubsection{Inductive Dataset Preparation}
In the inductive setting, we do not have access to all the users or items during the training process. New users/items can arrive post the model training, with new user-item interactions. Given these interactions, we need to derive embeddings for these new items/users without retraining. To evaluate in this setting, we hold out 5\% of users and items randomly (with a minimum of 10 and 5 interactions respectively) and remove them from the train, validation and test splits in the transductive setting. At the time of inference, we assume we have access to a partial set of interactions involving all the new users and new items using which we derive their embeddings. We then evaluate the model on the rest of the interactions, along with the test set.

\begin{table}[]
\centering
\resizebox{0.45\textwidth}{!}{%
\begin{tabular}{cccc}
\toprule
\textbf{Dataset}      & \textbf{User \#} & \textbf{Item \#} & \textbf{Interaction \#} \\ \midrule
\textbf{Gowalla}      & 29, 858          & 40, 981          & 1, 027, 370             \\
\textbf{Yelp2018}     & 31, 668          & 38, 048          & 1, 561, 406             \\
\textbf{Amazon-Book}  & 52, 643          & 91, 599          & 2, 984, 108             \\
\textbf{Douban-movie} & 3, 022           & 6, 971           & 195, 472    \\ \bottomrule          
\end{tabular}%
}
\caption{Statistics of benchmark datasets}
\label{tab:datasetStats}
\vspace{-10mm}
\end{table}
\vspace{-1mm}
\subsubsection{Methods of comparison} \lgcn gives a state-of-the-art performance. Hence \lgcn forms the primary baseline against which we compare all our proposed models. Additionally, we have included the following baselines.

\noindent\textbf{MF}~\cite{MF}: This is the traditional matrix factorization model that does not utilize the graph information directly.\\
\noindent\textbf{NGCF}~\cite{NeuralGraphCollaborativeFiltering}: Neural Graph Collaborative Filtering is a graph neural network based model that captures high-order information by embedding propagation using graphs. We utilized the code from this repository \footnote{https://github.com/kuandeng/LightGCN} to obtain performance metrics. \\
\noindent\textbf{Mult-VAE}~\cite{multvae}: This is a collaborative filtering method based on variational autoencoder. We run the code released by the authors\footnote{https://github.com/dawenl/vae\_cf} after modifying it to run on the train/val/test splits of our other experiments. \\
\noindent\textbf{GRMF-Norm}~\cite{GRMF}:  Graph Regularized Matrix Factroiation additonally adds a graph Laplacian regularizer in addition to the MF objective. We used the GRMF-Norm variant as described in~\cite{LGCN}\\
\noindent\textbf{GCN}: This baseline is equivalent to the non-linear version of LightGCN with transformation matrices learnt as is done in the traditional GCN layers.\\
\vspace{-5mm}
\subsubsection{Training Details}
We implemented all the above methods except NGCF and Mult-VAE. The training pipeline is setup identical to that of \lgcn for a fair comparison. We use BPR ~\cite{BPR} based loss function to train all the models using Adam Optimizer~\cite{adam}. All the models were trained for at most 1000 epochs with validation recall@20 evaluated every 20 epochs. There is early stopping when there is no improvement in the metric for ten consecutive evaluations. Additionally, we conducted all our experiments in a distributed setting, training on 4 GPU nodes using Horovod~\cite{horovod}. All the models we implemented are PyTorch based. We fixed embedding size to 64 for all the models. And the batch sizes for all datasets were fixed to 2048 except for Amazon-Book for which was set to 8192. Learning rate was swept over \{1e-1, 1e-2, 5e-3, 1e-3, 5e-4\}, embedding regularization over \{1e-2, 1e-3, 1e-4, 1e-5, 1e-6\}, graph dropout over \{0.0, 0.1, 0.2, 0.25\}. Additionally for NGCF, node dropout was swept over \{0.0, 0.1, 0.2, 0.25\} and the coefficient for graph laplacian regularizer for GRMF-Norm was swept over \{1e-2, 1e-3, 1e-4, 1e-5, 1e-6\}. We tuned the hyperparameters dropout rate in the range $[0, 0.2, 0.5]$, and $\beta$ in $[0.2, 0.4, 0.6, 0.8]$ for Mult-VAE with the model architecture as $200\rightarrow600\rightarrow200$, and the learning rate as 1e-3. For the proposed model, we also varied graph normalization as hyperparameters.

% \begin{enumerate}
%     \item Brief outline of the section 
%     \item Datasets description with table having statistics: \textbf{Table} 
%     \item Brief description of baselines
%     \item Transductive and Inductive Settings 
%     \item Hyper-parameter optimization and Metrics  
% \end{enumerate}

\vspace{-1mm}
\subsection{Transductive Setting Experiments} 
We carried out a detailed experimental study with the proposed CF-LGCN variants. We show the usefulness of twin networks first. Then, we present results with multiple layers and layer combination. We compare the recommended variants (\cflu) of our proposed approach to the state-of-the-art baselines.

% \subsubsection{CF-LGCN-U vs Twin CF-LGCN-U}
% \begin{table*}[t]
% \resizebox{0.75\textwidth}{!}{
% \begin{tabular}{|c|c|c|c|c|}
% \hline
% \textbf{Recall / NDCG @ 20} & \textbf{Gowalla} & \textbf{Yelp2018} & \textbf{Amazon-Book} & \textbf{Douban-Movie} \\ \hline
% \textbf{Twin CF-LGCN-U}     & 17.02 / 14.45    & 6.36 / 5.24       & \textbf{4.52 / 3.59} & 5.23 / 3.28           \\ \hline
% \textbf{CF-LGCN-U}          & 16.81 / 14.35    & 6.41 / 5.30       & 4.48 / 3.51          & 4.85 / 3.51           \\ \hline
% \end{tabular}
% }
% \caption{Twin CF-LGCN-U vs CF-LGCN-U}
% \label{tab:TableTwinLGCNvsLGCNU}
% \end{table*}
\vspace{-1mm}
\subsubsection{Single \cflu network Versus Twin \cflu network} In Table~\ref{tab:TableTwinLGCNvsLGCNU}, we present results obtained from evaluating single \cflu network and twin \cflu network on four benchmark datasets. We used three layers in each network. Recall that we learn two sets of user embedding with the twin network. We see that the twin \cflu network performs better than the single \cflu network on all datasets except \textsc{Yelp2018}. The results show that the twin \cflu network is beneficial, and can achieve higher performances by increasing the expressive power of \cflu network. Hence, we use only the twin network in the rest of our experiments. Performance on \textsc{Yelp2018} suggests that even a single network is sufficient in some application scenario. Therefore, it helps to experiment with both models and perform model section using the validation set. 

\begin{table}[]
\centering
\resizebox{0.49\textwidth}{!}{%
\begin{tabular}{@{}ccccc@{}}
\toprule
\textbf{Recall / NDCG @ 20} & \textbf{Gowalla}       & \textbf{Yelp2018}    & \textbf{Amazon-Book} & \textbf{Douban-Movie} \\ \midrule

CF-LGCN-U          & 16.81 / 14.35          & \textbf{6.41 / 5.30} & 4.48 / 3.51          & 4.85 / \textbf{3.51}           \\
Twin CF-LGCN-U     & \textbf{17.02 / 14.45} & 6.36 / 5.24          & \textbf{4.52 / 3.59} & \textbf{5.23} / 3.28  \\\bottomrule
\end{tabular}%
}
\caption{Twin CF-LGCN-U vs CF-LGCN-U}
\label{tab:TableTwinLGCNvsLGCNU}
\vspace{-10mm}
\end{table}
\vspace{-1mm}
\subsubsection{Experiment with Multiple Layers} In this experiment, we study the usefulness of having multiple layers in twin \cflu network. From Table~\ref{tab:MainTable}, we see that increasing the number of layers improves the performance significantly on \textsc{Yelp2018} and \textsc{Amazon-Book} datasets. This result demonstrates the usefulness of higher-order information available through propagation. The performance starts saturating, for example, in the \textsc{Gowalla} dataset. Besides, it is useful to note that increasing the number of layers beyond a certain limit may hurt the performance due to over-smoothing, as reported in node classification tasks \cite{GCN}. Therefore, it is crucial to treat layer size as a model hyperparameter and tune using validation set performance.      

\begin{figure}%
    \centering
    \subfigure{
        \label{fig:Amazon_metric}%
        \includegraphics[width=0.22\textwidth]{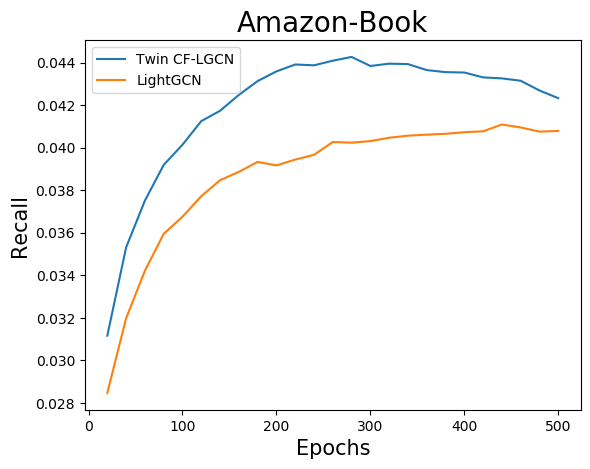}
    }
    \subfigure{
        \label{fig:Yelp_metric}
        \includegraphics[width=0.22\textwidth]{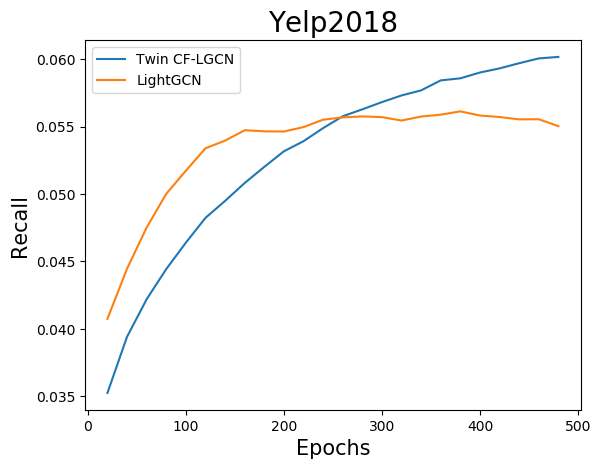}
    }\\
    \subfigure{
        \label{fig:Amazon_loss}%
        \includegraphics[width=0.225\textwidth]{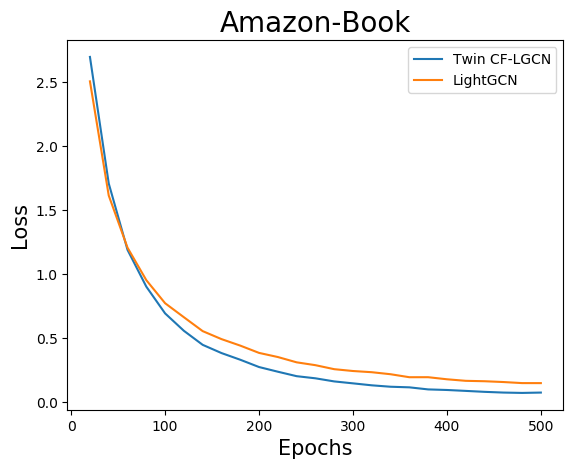}
    }
     \subfigure{
        \label{fig:Yelp_loss}
        \includegraphics[width=0.225\textwidth]{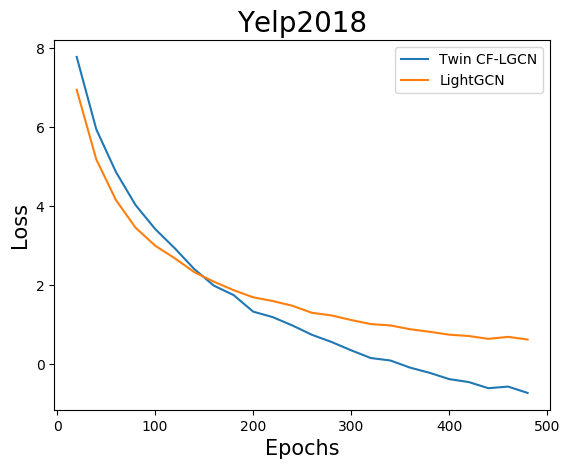}
    }
    \vspace{-7mm}
    \caption{Training loss and Test Recall @20 metric as as function of epochs}
    \label{fig:training_plots}
    \vspace{-3mm}
\end{figure}

% \begin{table*}[]
% \centering
% \resizebox{0.75\textwidth}{!}{%
% \begin{tabular}{@{}ccccc@{}}
% \toprule
% \textbf{Recall / NDCG @ 20}        & \textbf{Gowalla} & \textbf{Yelp2018}    & \textbf{Amazon-Book} & \textbf{Douban-Movie} \\ \midrule
% 1 Layer  & 17.08 / 14.48          & 6.20 / 5.13 & 3.87 / 3.07 & 4.91 / 3.19 \\
% 2 Layers & \textbf{17.09 / 14.51} & 6.31 / 5.19 & 4.32 / 3.44 & 5.17 / 3.23 \\
% 3 Layers & 17.02 / 14.45    & \textbf{6.36 / 5.24} & \textbf{4.52 / 3.59} & \textbf{5.23 / 3.28}                            \\ \bottomrule
% \end{tabular}%
% }
% \caption{Twin CF-LGCN-U Multi Layer Variants}
% \label{tab:MultiLayerVariants}
% \end{table*}

\subsubsection{Effect of Fusion} 
\lgcn reported that fusing embedding output from different layers helps to get improved performance compared to using only the final layer embedding output, and we observed a similar phenomenon in our experiments. We tried two variants of fusion: \textsc{Mean} and \textsc{Concat}. Following \lgcn, we used uniform weighting with \textsc{Mean}. From Table~\ref{tab:layer_fusion}, we see that \textsc{Concat} fusion delivers significant improvement with our twin \cflu network model. Therefore, we use \textsc{Concat} fusion with our model in rest of the experiments. Note that fusion can be done in several other ways, particularly when working with two networks. We leave this exercise for future experimental studies.

\subsubsection{Comparison with Baselines}
We performed a detailed experimental study by comparing the proposed model with several state-of-the-art baselines.In Figure~\ref{fig:training_plots}, we compare the training loss and recall$@$20 of the proposed twin \cflu model with \lgcn on the \textsc{Amazon-Book} and \textsc{Yelp2018} datasets. Detailed results are presented in Table~\ref{tab:MainTable}. We observe that the 3-layers twin \cflu network model performs uniformly better than the strongest competitor, \lgcn, except on the \textsc{Gowalla} dataset where the performance is very close. Thus, the proposed model is quite competitive to \lgcn and is a powerful alternative to \lgcn. As explained earlier, the \cflu network has the advantage of learning only user embedding. From Table~\ref{tab:datasetStats}, we see that the number of items is higher than the number of users, and is nearly twice for the \textsc{Amazon-Book} and \textsc{Douban-Movie} datasets. Thus, the proposed \cflu network can deliver similar or better performance than \lgcn with significantly reduced complexity, enabling large scale learning.     

\begin{table}[]
\centering
\resizebox{0.49\textwidth}{!}{%
\begin{tabular}{@{}cccccc@{}}
\toprule
\textbf{Recall / NDCG @ 20} & \textbf{Aggregation} & \textbf{Gowalla}       & \textbf{Yelp2018}    & \textbf{Amazon-Book} & \textbf{Douban-Movie} \\ \midrule
Twin CF-LGCN-U (2L) & Mean & 16.38 / 13.76 & \textbf{6.48 / 5.35} & 4.17 / 3.31 & 5.15 / 3.22 \\
Twin CF-LGCN-U (2L)         & Concat               & \textbf{17.09 / 14.51} & 6.31 / 5.19          & \textbf{4.32 / 3.44} & \textbf{5.17 / 3.23}  \\\midrule
Twin CF-LGCN-U (3L) & Mean & 16.71 / 14.27 & 6.06 / 5.03          & 4.24 / 3.31 & 5.16 / 3.17 \\
Twin CF-LGCN-U (3L)         & Concat               & \textbf{17.02 / 14.45} & \textbf{6.36 / 5.24} & \textbf{4.52 / 3.59} & \textbf{5.23 / 3.28}  \\ \bottomrule
\end{tabular}%
}
\caption{Different Layerwise Fusion}
\label{tab:layer_fusion}
\vspace{-10mm}
\end{table}

\begin{table*}[]
\centering
\resizebox{0.75\textwidth}{!}{%
\begin{tabular}{@{}ccccc@{}}
\toprule
\textbf{Recall / NDCG @ 20}  & \textbf{Gowalla}       & \textbf{Yelp2018}    & \textbf{Amazon-Book} & \textbf{Douban-Movie} \\ \midrule
\textbf{MF}                  & 14.55 / 11.56          & 5.17 / 4.16          & 3.49 / 2.67          & 4.24 / 2.77           \\
\textbf{NGCF}                & 15.69 / 12.82          & 5.60 / 4.55          & 3.85 / 2.94          & 4.85 / 3.00           \\
\textbf{Mult-VAE}            & 13.65 / 10.10          & 5.74 / 4.40          & 3.94 / 2.97          & 4.85 / 3.00           \\
\textbf{GRMF-Norm}           & 15.65 / 13.12          & 5.60 / 4.61          & 3.50 / 2.71          & 4.78 / 3.00           \\
\textbf{GCN (3L)}        & 16.10 / 13.76          & 6.14 / 5.06          & 3.67 / 2.84          & 4.78 / 3.02           \\ \midrule
\textbf{LightGCN (1L)}       & 17.11 / 14.59          & 6.17 / 5.04          & 3.76 / 2.93          & 4.75 / 3.03           \\
\textbf{LightGCN (2L)}       & 16.70 / 14.27          & 6.12 / 4.97          & 3.99 / 3.08          & 4.71 / 2.98           \\
\textbf{LightGCN (3L)}       & \textbf{17.21 / 14.65} & 6.07 / 5.00          & 4.10 / 3.16          & 4.95 / 3.14           \\ \midrule
\textbf{Twin CF-LGCN-U (1L)}  & 17.08 / 14.48          & 6.20 / 5.13 & 3.87 / 3.07 & 4.91 / 3.19 \\
\textbf{Twin CF-LGCN-U (2L)} & 17.09 / 14.51          & 6.31 / 5.19          & 4.32 / 3.44          & 5.17 / 3.23           \\
\textbf{Twin CF-LGCN-U (3L)} & 17.02 / 14.45          & \textbf{6.36 / 5.24} & \textbf{4.52 / 3.59} & \textbf{5.23 / 3.28}  \\ \bottomrule
\end{tabular}%
}
\caption{Comparison with Baselines}
\label{tab:MainTable}
\end{table*}

\subsubsection{Twin CF-LGCN-U vs Twin CF-LGCN-E}
Our \cf network architecture supports another variant \cfle. Though our interest primarily lies in \cflu, we compare these two variants. In Table~\ref{tab:TwinUvsE}, we report results from this experiment. We see that the twin \cflu network outperforms its counter-part twin \cfle on all datasets except the \textsc{Douban-Movie} dataset where the performance is very close. Note that the number of model parameters is significantly higher in the \cfle network, as $n > m$. Nevertheless, the performance is inferior. There has always been this question of developing user-centric versus item-centric models \cite{Yehuda_MF_journal} and studying this problem using \cf models is beyond the scope of this work. We leave this study for the future.   

\begin{table}[]
\centering
\resizebox{0.49\textwidth}{!}{%
\begin{tabular}{@{}ccccc@{}}
\toprule
\textbf{Recall / NDCG @ 20} & \textbf{Gowalla}       & \textbf{Yelp2018}      & \textbf{Amazon-Book} & \textbf{Douban-Movie}  \\ \midrule
Twin CF-LGCN-U     & \textbf{17.02 / 14.45} & \textbf{6.36   / 5.24} & \textbf{4.52 / 3.59} & 5.23   / 3.28          \\
Twin CF-LGCN-E     & 15.36 / 12.60          & 6.16   / 5.03          & 4.27   / 3.36        & \textbf{5.28   / 3.39} \\ \bottomrule
\end{tabular}%
}
\caption{Twin CF-LGCN-U vs Twin CF-LGCN-E}
\vspace{-8mm}
\label{tab:TwinUvsE}
\end{table}

% \begin{enumerate}
%     \item HeteGCN multi-layer variants results: \textbf{Table}  
%     \item HeteGCN comparison with baselines and LightGCN: \textbf{Table}  
%     \item Training loss/validation performance \textbf{Plots} as a function of epoch and wall-clock time 
%     \item Ablation studies: Graph normalization and Fusion (Mean and Concat), effect of non-linearity and transformation (if needed): \textbf{Table}  
% \end{enumerate}

\vspace{-1mm}
\subsection{Inductive Setting Experiments}

\subsubsection{Generalization to New Items and New Users}
We carried out a detailed experimental study to evaluate the effectiveness of inductive variants of CF-LGCN-U and \lgcn. We report three sets of metrics for both these models.\\
\noindent\textbf{Inductive}: This is a standard inductive setup where several users and items are unseen during training. During inference, we use partial interactions available for all the new entities to obtain their embeddings. Note that we do not need to retrain these models. \\
\noindent\textbf{Transductive (Upper Bound)}: In this setup, we train the model along with the partial interactions, thereby having access to all the users and items during training. The motivation to report these numbers is to get an upper bound on the performance.\\
\noindent\textbf{Transductive (Lower Bound)}: This set of numbers indicates the lower bound and highlights the performance gain that one can achieve by recommending for new users and items. Training and inference are made exactly like the inductive setting. However, we assume that we cannot make recommendations for new users and new items during the evaluation. \\

From the results in Table \ref{tab:InductiveTable}, the inductive variants of both CF-LGCN-U and \lgcn obtain significant gains over the corresponding Transductive (Lower Bound) metrics and closer to the Transductive (Upper Bound) metrics. This result indicates that the proposed inductive modification is quite effective in generalizing to new users and items. CF-LGCN-U performs better than \lgcn on \textsc{Yelp2018} and \textsc{Amazon-Book} datasets, while it is marginally inferior in the \textsc{Gowalla} dataset (as was observed in the transductive setting).

\subsubsection{Effect of Updating User Embedding}
Additionally, we carried out a few experiments in a setting where we had access to all the users during training and evaluated the performance only on new items. As we have access to all the user embedding, we only need to derive embedding for new items. We can derive item embedding while retaining the user embedding as it is, or we could update the embedding for users as well when we get access to the partial interactions for the new items. We report metrics for both these cases in Table \ref{tab:Uemb_upd}. As we can see, updating user embedding with partially observed interactions during inference can be very useful over using static user embedding.

\begin{table}[]
\centering
\resizebox{0.5\textwidth}{!}{%
\begin{tabular}{@{}cccc@{}}
\toprule
\textbf{Recall / NDCG @ 20} & \textbf{Gowalla} & \textbf{Yelp2018}    & \textbf{Amazon-Book} \\ \toprule
\multicolumn{4}{c}{\textbf{Transductive (Lower Bound)}}                      \\ \midrule
LightGCN       & 15.58 / 12.60          & 5.79 / 4.87 & 3.66 / 2.94 \\
Twin CF-LGCN-U & 14.49 / 12.49          & 5.78 / 4.92 & 4.02 / 3.29 \\ \toprule
\multicolumn{4}{c}{\textbf{Inductive}}                                       \\ \midrule
LightGCN       & \textbf{16.79 / 14.80} & 6.38 / 5.58 & 4.28 / 3.69 \\
Twin CF-LGCN-U    & 16.71 / 14.68    & \textbf{6.44 / 5.69} & \textbf{4.88 / 4.26} \\ \toprule
\multicolumn{4}{c}{\textbf{Transductive (Upper Bound)}}                      \\ \midrule
LightGCN       & 17.19 / 15.34          & 6.44 / 5.65 & 4.59 / 3.98 \\
Twin CF-LGCN-U & 17.09 / 15.11          & 6.48 / 5.71 & 5.11 / 4.49 \\ \bottomrule
\end{tabular}%
}
\caption{Generalization to New Users and New Items}
\label{tab:InductiveTable}
\vspace{-3mm}
\end{table}

\vspace{-3mm}
% Please add the following required packages to your document preamble:
% \usepackage{booktabs}
% \usepackage{graphicx}
\begin{table}[]
\centering
\resizebox{0.5\textwidth}{!}{%
\begin{tabular}{@{}cccc@{}}
\toprule
\textbf{Recall / NDCG @ 20}  & \textbf{Gowalla}         & \textbf{Yelp2018}      & \textbf{Amazon-Book} \\ \midrule
Twin CF-LGCN-U (2L)          & 16.12 /   13.72          & 6.11 /   5.19          & 4.08 /   3.33        \\Twin CF-LGCN-U (2L)  with U+ & \textbf{16.68 /   14.45} & \textbf{6.23 /   5.29} & \textbf{4.23 / 3.48} \\
 \bottomrule
\end{tabular}%
}
\caption{U+ suffix represents the model for which we update user embedding during inference}
\label{tab:Uemb_upd}
\vspace{-10mm}
\end{table}

\vspace{3mm}
\section{Related Work}
% We cover the following areas. 

% \begin{enumerate}
%     \item Collaborative Filtering and Matrix Factorization related methods 
%     \item Graph based Recommender Systems 
%     \item Inductive Recommendation 
% \end{enumerate}

Recommendation problems are ubiquitous in our day-to-day life. User-Item interaction graphs capture historical information on users interaction with various items. Collaborative Filtering (CF) methods utilize these interaction graphs to make item recommendations to users. User-item interactions have several dimensions:\\
\textbf{1. Binary / Real Valued:} Interactions could be binary (e.g. user bought a product or not) or real-valued (e.g. user gave a 5/10 rating on a video). This distinction might seem trivial; however, there are several subtle differences between the two. In binary interaction graphs, we are often interested in predicting top-k items for a user; hence, the interest metric is often NDCG@k. In real-valued interactions like ratings, we are interested in predicting user's rating for an item (regardless of the user's preference towards that item). Thereby, the metric often measured here is the MSE. Also, there are inherent biases involved in real-valued interactions - a user may be prone to giving high ratings to all products, or a popular item is prone to receiving high ratings. Such biases are absent in a binary interaction graph.\\
\textbf{2. Explicit / Implicit:} Interactions could be explicit (e.g. user purchased a product) or implicit (e.g. user watched a YouTube video). Explicit feedbacks create a reliable user-interaction graph which can lead to a reliable recommendation. If a user purchases an item, we can be sure that they liked/sought the item. Hence, recommendations learnt from such graphs have a high degree of confidence in being liked/sought by the user. However, implicit feedback, due to its nature, creates a highly unreliable graph. Just because a user watched a YouTube video does not imply that the user liked it. Any recommendation made on such unreliable information can be noisy.

Several methods have been proposed for CF over the years, addressing different aspects of the recommendation problem.\\
% Recommendation of items like movies, music, products etc. have gained lot of relevance, particularly during these COVID times. The interaction of users with the items can be captured via a user-item interaction graph. These user interactions could be binary for e.g. user bought a product or not etc. or they could be real valued user gave a rating of 5/10 for a video etc. On the other hand, interactions could be explicit (for e.g user purchased a product) or implicit (for e.g. user watched a YouTube video). Collaborative Filtering (CF) methods utilize these interaction graphs to make item recommendations to users. 
\textbf{Factorization Methods:} Early methods assumed fixed user and item set, and proposed Matrix Factorization (MF) based methods~\cite{SVD_Golub, Yehuda_MF_recommender_systems, ALS_PR} which treats the user-item interaction graph as a matrix. These methods can successfully handle binary interactions. For real-valued interactions, addressing inherent biases is essential.  \cite{Yehuda_MF_journal} computes explicit per-user per-item bias estimates to get improved performance. In cases of implicit interactions, it is necessary to identify meaningful interactions for downstream recommendation problems; hence attention-based models have been developed for such scenarios that use side-information to model attention \cite{attentionCF, NAIS}. With the advent of deep learning, non-linear models of matrix factorization have been proposed \cite{NeuralCollaborativeFiltering} and models that can leverage rich side information to get improved recommendation \cite{CollaborativeDeepLearning, CollborativeKnowledgeBase}. AutoEncoder-based models~\cite{collabdenoise, multvae} have been proposed to address implicit interactions. \\
\textbf{GNN based:}Viewing graphs as matrices has certain inherent limitations. Since not all users interact with all items, certain entries in the matrix are missing. Replacing them with default values addresses these missing entry problems. Choice of the default value often dictates the performance of the model. Additionally, these methods do not explicitly exploit the collaborative nature of the user-item interaction graph. Researchers have started to look at Graph Neural Networks (GNNs)~\cite{survey_wu} to circumvent these issues. Graph Convolutional Network (GCN)~\cite{GCN}, a particular instantiation of GNNs, has shown to be capable of exploiting graphs to give improved performance. \cite{GCMC} adapt GCNs for user-item interaction graph and show that it can improve performance over baseline MF methods; however, it only uses single-hop information. \cite{NeuralGraphCollaborativeFiltering} adapts the model to incorporate multiple hop information. \cite{PinSage} developed a random walk based GCN model that can work for a massive graph. \cite{LGCN} proposed a simplified version of the GCN where neighborhood aggregation alone gave good gains. \\
\textbf{Inductive:} All the above methods, however, assume a fixed user-item set. This assumption can be very restrictive, particularly in social network and e-commerce scenarios where several new users and new items get added every day. In such cases, it is desirable to have an inductive method which does not assume a fixed user-item set. Our proposed approach is one such method. In this approach, we derive item embeddings from user-embeddings and user-embeddings can be derived from item-embeddings. Exploiting such relationship allows the model to generalize to new users and new items without retraining. Our proposed approach is competitive with existing approaches in transductive settings, showing that it is possible to express item-embeddings in terms of user-embeddings and vice versa. It also performs well in inductive settings without degradation, proving its generalizability to new users and new items. There have been a few prior works in inductive recommendation \cite{DeepModelsOfInteractions, IGMC}. However, these methods work on real-valued interaction graphs. As discussed earlier, the real-valued interactions have implicit biases. These are either addressed by either estimating per user, per item bias as done in~\cite{Yehuda_MF_journal} or modeling each real-valued possibility as a separate relation as done in \cite{IGMC}. Addressing these biases is a non-trivial problem and beyond the scope of the work. We refrain from comparing with these approaches as we restrict our focus to binary-valued user-item interaction graphs. 
\vspace{-1mm}
\section{Discussion and Future Work}
We focused our attention on applying the proposed modeling approach to scenarios where user-interaction data is available as binary information and user-specific recommendation is made, with recommendation quality measured using metrics such as recall$@k$ and ndcg$@k$.
Learning models in this application setting comes with several other important factors, including negative sampling~\cite{NegSamp} and choice of loss functions (e.g., approximate ndcg~\cite{ApproxNDCG}) that help to get improved performance. Incorporating these factors and conducting an experimental study is beyond the scope of this work.  

Another important recommendation application is \textit{rating prediction} task (e.g., users rating movies or products)~\cite{Yehuda_MF_recommender_systems}. Several methods have been developed and studied recently. This includes graph neural networks based matrix completion~\cite{GCMC, IGMC}, heterogeneous (graph) information networks~\cite{HINsForRecommendation} which uses concepts such as meta-path where a meta-path (e.g., User - Book - Author - Book) encodes semantic information with higher order relations. Further, several solutions to address \textit{cold-start} (e.g., recommending for users/items having only a few rating entries) and inductive inference~\cite{IGMC}. In some application setting, knowledge graphs (e.g., explicit user-user relational graphs) are available, and they are used to improve the recommendation quality. See~\cite{DL_on_KG_recommender_system} for more details.  

We can adapt our \cf modeling approach for these different application scenarios by modifying network architecture, for example, using GCN with additional graph types~\cite{HeteGCN}, relational GCN (R-GCN)~\cite{RGCN}, Reco-GCN~\cite{RecoGCN}) and optimizing for metrics such as mean-squared-error (MSE), mean-absolute error (MAE) and ndcg$@k$ etc. We leave all these directions for future work.

\vspace{-1mm}
\section{Conclusion}
We consider the problem of learning graph embedding based method for collaborative filtering (CF). We set out to develop an inductive recommendation model with graph neural networks and build scalable models that have reduced model complexity. We proposed a novel graph convolutional networks modeling approach for collaborating (\cf). Among the possible variants within this approach's scope, our primary candidate is the \cflu network and its twin variant. The \cflu network model learns only user embedding. This network offers inductive recommendation capability, therefore, generalizes for users and items unseen during training.
Furthermore, its model complexity is dependent \textit{only} on the number of users. Thus, its model complexity is significantly lesser compared to even light models such as \lgcn. \cflu models are quite attractive in many practical application scenarios, where the number of items is significantly lesser than the number of users. We showed the relation between the proposed models and \lgcn. Our experimental results demonstrate the efficacy of the proposed modeling approach in both transductive and inductive settings.  

%% the bibliography file.
\bibliographystyle{ACM-Reference-Format}
\bibliography{references} %%bibliography

%%% -*-BibTeX-*-
%%% Do NOT edit. File created by BibTeX with style
%%% ACM-Reference-Format-Journals [18-Jan-2012].

\begin{thebibliography}{47}

%%% ====================================================================
%%% NOTE TO THE USER: you can override these defaults by providing
%%% customized versions of any of these macros before the \bibliography
%%% command.  Each of them MUST provide its own final punctuation,
%%% except for \shownote{}, \showDOI{}, and \showURL{}.  The latter two
%%% do not use final punctuation, in order to avoid confusing it with
%%% the Web address.
%%%
%%% To suppress output of a particular field, define its macro to expand
%%% to an empty string, or better, \unskip, like this:
%%%
%%% \newcommand{\showDOI}[1]{\unskip}   % LaTeX syntax
%%%
%%% \def \showDOI #1{\unskip}           % plain TeX syntax
%%%
%%% ====================================================================

\ifx \showCODEN    \undefined \def \showCODEN     #1{\unskip}     \fi
\ifx \showDOI      \undefined \def \showDOI       #1{#1}\fi
\ifx \showISBNx    \undefined \def \showISBNx     #1{\unskip}     \fi
\ifx \showISBNxiii \undefined \def \showISBNxiii  #1{\unskip}     \fi
\ifx \showISSN     \undefined \def \showISSN      #1{\unskip}     \fi
\ifx \showLCCN     \undefined \def \showLCCN      #1{\unskip}     \fi
\ifx \shownote     \undefined \def \shownote      #1{#1}          \fi
\ifx \showarticletitle \undefined \def \showarticletitle #1{#1}   \fi
\ifx \showURL      \undefined \def \showURL       {\relax}        \fi
% The following commands are used for tagged output and should be
% invisible to TeX
\providecommand\bibfield[2]{#2}
\providecommand\bibinfo[2]{#2}
\providecommand\natexlab[1]{#1}
\providecommand\showeprint[2][]{arXiv:#2}

\bibitem[\protect\citeauthoryear{Bruch, Zoghi, Bendersky, and Najork}{Bruch
  et~al\mbox{.}}{2019}]%
        {ApproxNDCG}
\bibfield{author}{\bibinfo{person}{Sebastian Bruch}, \bibinfo{person}{Masrour
  Zoghi}, \bibinfo{person}{Michael Bendersky}, {and} \bibinfo{person}{Marc
  Najork}.} \bibinfo{year}{2019}\natexlab{}.
\newblock \showarticletitle{Revisiting Approximate Metric Optimization in the
  Age of Deep Neural Networks}. In \bibinfo{booktitle}{\emph{Proceedings of the
  42nd International ACM SIGIR Conference on Research and Development in
  Information Retrieval}} (Paris, France) \emph{(\bibinfo{series}{SIGIR'19})}.
  \bibinfo{publisher}{Association for Computing Machinery},
  \bibinfo{address}{New York, NY, USA}, \bibinfo{pages}{1241–1244}.
\newblock
\showISBNx{9781450361729}


\bibitem[\protect\citeauthoryear{Cao, Lin, Guo, Liu, Liu, and Wang}{Cao
  et~al\mbox{.}}{2021}]%
        {BGE}
\bibfield{author}{\bibinfo{person}{Jiangxia Cao}, \bibinfo{person}{Xixun Lin},
  \bibinfo{person}{Shu Guo}, \bibinfo{person}{Luchen Liu},
  \bibinfo{person}{Tingwen Liu}, {and} \bibinfo{person}{Bin Wang}.}
  \bibinfo{year}{2021}\natexlab{}.
\newblock \showarticletitle{Bipartite Graph Embedding via Mutual Information
  Maximization}. In \bibinfo{booktitle}{\emph{WSDM}}.
\newblock


\bibitem[\protect\citeauthoryear{Chen, Zhang, He, Nie, Liu, and Chua}{Chen
  et~al\mbox{.}}{2017}]%
        {attentionCF}
\bibfield{author}{\bibinfo{person}{Jingyuan Chen}, \bibinfo{person}{Hanwang
  Zhang}, \bibinfo{person}{Xiangnan He}, \bibinfo{person}{Liqiang Nie},
  \bibinfo{person}{Wei Liu}, {and} \bibinfo{person}{Tat-Seng Chua}.}
  \bibinfo{year}{2017}\natexlab{}.
\newblock \showarticletitle{Attentive Collaborative Filtering: Multimedia
  Recommendation with Item- and Component-Level Attention}. In
  \bibinfo{booktitle}{\emph{Proceedings of the 40th International ACM SIGIR
  Conference on Research and Development in Information Retrieval}}.
\newblock


\bibitem[\protect\citeauthoryear{{Chen}, {Hua}, {Chang}, {Wang}, {Zhang}, and
  {Kong}}{{Chen} et~al\mbox{.}}{2018}]%
        {CF_Survey1}
\bibfield{author}{\bibinfo{person}{R. {Chen}}, \bibinfo{person}{Q. {Hua}},
  \bibinfo{person}{Y. {Chang}}, \bibinfo{person}{B. {Wang}},
  \bibinfo{person}{L. {Zhang}}, {and} \bibinfo{person}{X. {Kong}}.}
  \bibinfo{year}{2018}\natexlab{}.
\newblock \showarticletitle{A Survey of Collaborative Filtering-Based
  Recommender Systems: From Traditional Methods to Hybrid Methods Based on
  Social Networks}.
\newblock \bibinfo{journal}{\emph{IEEE Access}}  \bibinfo{volume}{6}
  (\bibinfo{year}{2018}), \bibinfo{pages}{64301--64320}.
\newblock


\bibitem[\protect\citeauthoryear{Dziugaite and Roy}{Dziugaite and Roy}{2015}]%
        {NMF}
\bibfield{author}{\bibinfo{person}{Gintare~Karolina Dziugaite} {and}
  \bibinfo{person}{Daniel~M. Roy}.} \bibinfo{year}{2015}\natexlab{}.
\newblock \showarticletitle{Neural Network Matrix Factorization}.
\newblock \bibinfo{journal}{\emph{CoRR}}  \bibinfo{volume}{abs/1511.06443}
  (\bibinfo{year}{2015}).
\newblock
\showeprint[arxiv]{1511.06443}


\bibitem[\protect\citeauthoryear{Ekstrand, Riedl, and Konstan}{Ekstrand
  et~al\mbox{.}}{2011}]%
        {CF_RecSys}
\bibfield{author}{\bibinfo{person}{Michael~D. Ekstrand},
  \bibinfo{person}{John~T. Riedl}, {and} \bibinfo{person}{Joseph~A. Konstan}.}
  \bibinfo{year}{2011}\natexlab{}.
\newblock \showarticletitle{Collaborative Filtering Recommender Systems}.
\newblock \bibinfo{journal}{\emph{Found. Trends Hum.-Comput. Interact.}}
  \bibinfo{volume}{4}, \bibinfo{number}{2} (\bibinfo{date}{Feb.}
  \bibinfo{year}{2011}), \bibinfo{pages}{81–173}.
\newblock
\showISSN{1551-3955}


\bibitem[\protect\citeauthoryear{Gao, Li, Lin, Gao, and Khan}{Gao
  et~al\mbox{.}}{2020}]%
        {DL_on_KG_recommender_system}
\bibfield{author}{\bibinfo{person}{Yang Gao}, \bibinfo{person}{Yi-Fan Li},
  \bibinfo{person}{Yu Lin}, \bibinfo{person}{Hang Gao}, {and}
  \bibinfo{person}{Latifur Khan}.} \bibinfo{year}{2020}\natexlab{}.
\newblock \bibinfo{title}{Deep Learning on Knowledge Graph for Recommender
  System: A Survey}.
\newblock
\newblock
\showeprint[arxiv]{2004.00387}~[cs.IR]


\bibitem[\protect\citeauthoryear{{George} and {Merugu}}{{George} and
  {Merugu}}{2005}]%
        {scalablecf_sru}
\bibfield{author}{\bibinfo{person}{T. {George}} {and} \bibinfo{person}{S.
  {Merugu}}.} \bibinfo{year}{2005}\natexlab{}.
\newblock \showarticletitle{A scalable collaborative filtering framework based
  on co-clustering}. In \bibinfo{booktitle}{\emph{Fifth IEEE International
  Conference on Data Mining (ICDM'05)}}.
\newblock


\bibitem[\protect\citeauthoryear{Golub and Reinsch}{Golub and Reinsch}{1970}]%
        {SVD_Golub}
\bibfield{author}{\bibinfo{person}{G.~H. Golub} {and} \bibinfo{person}{C.
  Reinsch}.} \bibinfo{year}{1970}\natexlab{}.
\newblock \showarticletitle{Singular Value Decomposition and Least Squares
  Solutions}.
\newblock \bibinfo{journal}{\emph{Numer. Math.}} \bibinfo{volume}{14},
  \bibinfo{number}{5} (\bibinfo{date}{April} \bibinfo{year}{1970}),
  \bibinfo{pages}{403–420}.
\newblock
\showISSN{0029-599X}


\bibitem[\protect\citeauthoryear{Hartford, Graham, Leyton-Brown, and
  Ravanbakhsh}{Hartford et~al\mbox{.}}{2018}]%
        {DeepModelsOfInteractions}
\bibfield{author}{\bibinfo{person}{Jason Hartford}, \bibinfo{person}{Devon
  Graham}, \bibinfo{person}{Kevin Leyton-Brown}, {and} \bibinfo{person}{Siamak
  Ravanbakhsh}.} \bibinfo{year}{2018}\natexlab{}.
\newblock \showarticletitle{Deep Models of Interactions Across Sets}. In
  \bibinfo{booktitle}{\emph{Proceedings of the 35th International Conference on
  Machine Learning}} \emph{(\bibinfo{series}{Proceedings of Machine Learning
  Research}, Vol.~\bibinfo{volume}{80})},
  \bibfield{editor}{\bibinfo{person}{Jennifer Dy} {and}
  \bibinfo{person}{Andreas Krause}} (Eds.). \bibinfo{publisher}{PMLR},
  \bibinfo{address}{Stockholmsmässan, Stockholm Sweden},
  \bibinfo{pages}{1909--1918}.
\newblock


\bibitem[\protect\citeauthoryear{He and McAuley}{He and McAuley}{2016}]%
        {Amazon}
\bibfield{author}{\bibinfo{person}{Ruining He} {and} \bibinfo{person}{Julian~J.
  McAuley}.} \bibinfo{year}{2016}\natexlab{}.
\newblock \showarticletitle{Ups and Downs: Modeling the Visual Evolution of
  Fashion Trends with One-Class Collaborative Filtering}.
\newblock  (\bibinfo{year}{2016}).
\newblock


\bibitem[\protect\citeauthoryear{He and Chua}{He and Chua}{2017}]%
        {Neural_Factorization_Machine_Predictive_Sparse_Analytics}
\bibfield{author}{\bibinfo{person}{Xiangnan He} {and} \bibinfo{person}{Tat-Seng
  Chua}.} \bibinfo{year}{2017}\natexlab{}.
\newblock \showarticletitle{Neural Factorization Machines for Sparse Predictive
  Analytics}. In \bibinfo{booktitle}{\emph{Proceedings of the 40th
  International ACM SIGIR Conference on Research and Development in Information
  Retrieval}} (Shinjuku, Tokyo, Japan) \emph{(\bibinfo{series}{SIGIR '17})}.
  \bibinfo{publisher}{Association for Computing Machinery},
  \bibinfo{address}{New York, NY, USA}.
\newblock
\showISBNx{9781450350228}


\bibitem[\protect\citeauthoryear{He, Deng, Wang, Li, Zhang, and Wang}{He
  et~al\mbox{.}}{2020}]%
        {LGCN}
\bibfield{author}{\bibinfo{person}{Xiangnan He}, \bibinfo{person}{Kuan Deng},
  \bibinfo{person}{Xiang Wang}, \bibinfo{person}{Yan Li},
  \bibinfo{person}{YongDong Zhang}, {and} \bibinfo{person}{Meng Wang}.}
  \bibinfo{year}{2020}\natexlab{}.
\newblock \bibinfo{booktitle}{\emph{LightGCN: Simplifying and Powering Graph
  Convolution Network for Recommendation}}.
\newblock \bibinfo{publisher}{Association for Computing Machinery},
  \bibinfo{address}{New York, NY, USA}, \bibinfo{pages}{639–648}.
\newblock
\showISBNx{9781450380164}


\bibitem[\protect\citeauthoryear{{He}, {He}, {Song}, {Liu}, {Jiang}, and
  {Chua}}{{He} et~al\mbox{.}}{2018}]%
        {NAIS}
\bibfield{author}{\bibinfo{person}{X. {He}}, \bibinfo{person}{Z. {He}},
  \bibinfo{person}{J. {Song}}, \bibinfo{person}{Z. {Liu}}, \bibinfo{person}{Y.
  {Jiang}}, {and} \bibinfo{person}{T. {Chua}}.}
  \bibinfo{year}{2018}\natexlab{}.
\newblock \showarticletitle{NAIS: Neural Attentive Item Similarity Model for
  Recommendation}.
\newblock \bibinfo{journal}{\emph{IEEE Transactions on Knowledge and Data
  Engineering}} \bibinfo{volume}{30}, \bibinfo{number}{12}
  (\bibinfo{year}{2018}), \bibinfo{pages}{2354--2366}.
\newblock


\bibitem[\protect\citeauthoryear{He, Liao, Zhang, Nie, Hu, and Chua}{He
  et~al\mbox{.}}{2017}]%
        {NeuralCollaborativeFiltering}
\bibfield{author}{\bibinfo{person}{Xiangnan He}, \bibinfo{person}{Lizi Liao},
  \bibinfo{person}{Hanwang Zhang}, \bibinfo{person}{Liqiang Nie},
  \bibinfo{person}{Xia Hu}, {and} \bibinfo{person}{Tat-Seng Chua}.}
  \bibinfo{year}{2017}\natexlab{}.
\newblock \showarticletitle{Neural Collaborative Filtering}
  \emph{(\bibinfo{series}{WWW '17})}. \bibinfo{publisher}{International World
  Wide Web Conferences Steering Committee}, \bibinfo{address}{Republic and
  Canton of Geneva, CHE}.
\newblock
\showISBNx{9781450349130}


\bibitem[\protect\citeauthoryear{Kabbur, Ning, and Karypis}{Kabbur
  et~al\mbox{.}}{2013}]%
        {fism}
\bibfield{author}{\bibinfo{person}{Santosh Kabbur}, \bibinfo{person}{Xia Ning},
  {and} \bibinfo{person}{George Karypis}.} \bibinfo{year}{2013}\natexlab{}.
\newblock \showarticletitle{FISM: Factored Item Similarity Models for Top-N
  Recommender Systems}. In \bibinfo{booktitle}{\emph{Proceedings of the 19th
  ACM SIGKDD}}.
\newblock


\bibitem[\protect\citeauthoryear{Karydi and Margaritis}{Karydi and
  Margaritis}{2016}]%
        {pardisCF}
\bibfield{author}{\bibinfo{person}{Efthalia Karydi} {and}
  \bibinfo{person}{Konstantinos Margaritis}.} \bibinfo{year}{2016}\natexlab{}.
\newblock \showarticletitle{Parallel and Distributed Collaborative Filtering: A
  Survey}.
\newblock \bibinfo{journal}{\emph{ACM Comput. Surv.}} \bibinfo{volume}{49},
  \bibinfo{number}{2}, Article \bibinfo{articleno}{37} (\bibinfo{date}{Aug.}
  \bibinfo{year}{2016}), \bibinfo{numpages}{41}~pages.
\newblock


\bibitem[\protect\citeauthoryear{Kingma and Ba}{Kingma and Ba}{2015}]%
        {adam}
\bibfield{author}{\bibinfo{person}{Diederik~P. Kingma} {and}
  \bibinfo{person}{Jimmy Ba}.} \bibinfo{year}{2015}\natexlab{}.
\newblock \showarticletitle{Adam: A Method for Stochastic Optimization}. In
  \bibinfo{booktitle}{\emph{ICLR}}.
\newblock


\bibitem[\protect\citeauthoryear{Kipf and Welling}{Kipf and Welling}{2017}]%
        {GCN}
\bibfield{author}{\bibinfo{person}{Thomas~N. Kipf} {and} \bibinfo{person}{Max
  Welling}.} \bibinfo{year}{2017}\natexlab{}.
\newblock \showarticletitle{Semi-Supervised Classification with Graph
  Convolutional Networks}. In \bibinfo{booktitle}{\emph{ICLR}}.
\newblock


\bibitem[\protect\citeauthoryear{Koren}{Koren}{2008}]%
        {Yehuda_MF_journal}
\bibfield{author}{\bibinfo{person}{Yehuda Koren}.}
  \bibinfo{year}{2008}\natexlab{}.
\newblock \showarticletitle{Factorization Meets the Neighborhood: A
  Multifaceted Collaborative Filtering Model}. In
  \bibinfo{booktitle}{\emph{Proceedings of the 14th ACM SIGKDD}}.
  \bibinfo{publisher}{Association for Computing Machinery},
  \bibinfo{address}{New York, NY, USA}.
\newblock
\showISBNx{9781605581934}


\bibitem[\protect\citeauthoryear{{Koren}, {Bell}, and {Volinsky}}{{Koren}
  et~al\mbox{.}}{2009}]%
        {Yehuda_MF_recommender_systems}
\bibfield{author}{\bibinfo{person}{Y. {Koren}}, \bibinfo{person}{R. {Bell}},
  {and} \bibinfo{person}{C. {Volinsky}}.} \bibinfo{year}{2009}\natexlab{}.
\newblock \showarticletitle{Matrix Factorization Techniques for Recommender
  Systems}.
\newblock \bibinfo{journal}{\emph{Computer}} \bibinfo{volume}{42},
  \bibinfo{number}{8} (\bibinfo{year}{2009}), \bibinfo{pages}{30--37}.
\newblock


\bibitem[\protect\citeauthoryear{Liang, Charlin, McInerney, and Blei}{Liang
  et~al\mbox{.}}{2016}]%
        {gowalla}
\bibfield{author}{\bibinfo{person}{Dawen Liang}, \bibinfo{person}{Laurent
  Charlin}, \bibinfo{person}{James McInerney}, {and} \bibinfo{person}{David~M.
  Blei}.} \bibinfo{year}{2016}\natexlab{}.
\newblock \showarticletitle{Modeling User Exposure in Recommendation}. In
  \bibinfo{booktitle}{\emph{Proceedings of the 25th International Conference on
  World Wide Web}} (Montr\'{e}al, Qu\'{e}bec, Canada)
  \emph{(\bibinfo{series}{WWW '16})}. \bibinfo{publisher}{International World
  Wide Web Conferences Steering Committee}, \bibinfo{address}{Republic and
  Canton of Geneva, CHE}, \bibinfo{pages}{951–961}.
\newblock
\showISBNx{9781450341431}


\bibitem[\protect\citeauthoryear{Liang, Krishnan, Hoffman, and Jebara}{Liang
  et~al\mbox{.}}{2018}]%
        {multvae}
\bibfield{author}{\bibinfo{person}{Dawen Liang}, \bibinfo{person}{Rahul~G.
  Krishnan}, \bibinfo{person}{Matthew~D. Hoffman}, {and} \bibinfo{person}{Tony
  Jebara}.} \bibinfo{year}{2018}\natexlab{}.
\newblock \showarticletitle{Variational Autoencoders for Collaborative
  Filtering}. In \bibinfo{booktitle}{\emph{Proceedings of the 2018 World Wide
  Web Conference}}.
\newblock


\bibitem[\protect\citeauthoryear{{Papadimitriou} and {Sun}}{{Papadimitriou} and
  {Sun}}{2008}]%
        {disco}
\bibfield{author}{\bibinfo{person}{S. {Papadimitriou}} {and}
  \bibinfo{person}{J. {Sun}}.} \bibinfo{year}{2008}\natexlab{}.
\newblock \showarticletitle{DisCo: Distributed Co-clustering with Map-Reduce: A
  Case Study towards Petabyte-Scale End-to-End Mining}. In
  \bibinfo{booktitle}{\emph{2008 Eighth IEEE International Conference on Data
  Mining}}.
\newblock


\bibitem[\protect\citeauthoryear{Peng and Mine}{Peng and Mine}{2020}]%
        {HierGCN}
\bibfield{author}{\bibinfo{person}{Shaowen Peng} {and}
  \bibinfo{person}{Tsunenori Mine}.} \bibinfo{year}{2020}\natexlab{}.
\newblock \bibinfo{title}{A Robust Hierarchical Graph Convolutional Network
  Model for Collaborative Filtering}.
\newblock
\newblock
\showeprint[arxiv]{2004.14734}~[cs.IR]


\bibitem[\protect\citeauthoryear{Ragesh, Sellamanickam, Iyer, Bairi, and
  Lingam}{Ragesh et~al\mbox{.}}{2021}]%
        {HeteGCN}
\bibfield{author}{\bibinfo{person}{Rahul Ragesh}, \bibinfo{person}{Sundararajan
  Sellamanickam}, \bibinfo{person}{Arun Iyer}, \bibinfo{person}{Ram Bairi},
  {and} \bibinfo{person}{Vijay Lingam}.} \bibinfo{year}{2021}\natexlab{}.
\newblock \showarticletitle{HeteGCN: Heterogeneous Graph Convolutional Networks
  for Text Classification}. In \bibinfo{booktitle}{\emph{WSDM}}.
\newblock


\bibitem[\protect\citeauthoryear{Rao, Yu, Ravikumar, and Dhillon}{Rao
  et~al\mbox{.}}{2015}]%
        {GRMF}
\bibfield{author}{\bibinfo{person}{Nikhil Rao}, \bibinfo{person}{Hsiang{-}Fu
  Yu}, \bibinfo{person}{Pradeep Ravikumar}, {and} \bibinfo{person}{Inderjit~S.
  Dhillon}.} \bibinfo{year}{2015}\natexlab{}.
\newblock \showarticletitle{Collaborative Filtering with Graph Information:
  Consistency and Scalable Methods}. In \bibinfo{booktitle}{\emph{Advances in
  Neural Information Processing Systems 28: Annual Conference on Neural
  Information Processing Systems 2015, December 7-12, 2015, Montreal, Quebec,
  Canada}}, \bibfield{editor}{\bibinfo{person}{Corinna Cortes},
  \bibinfo{person}{Neil~D. Lawrence}, \bibinfo{person}{Daniel~D. Lee},
  \bibinfo{person}{Masashi Sugiyama}, {and} \bibinfo{person}{Roman Garnett}}
  (Eds.). \bibinfo{pages}{2107--2115}.
\newblock


\bibitem[\protect\citeauthoryear{{Rendle}}{{Rendle}}{2010}]%
        {Factorization_machines}
\bibfield{author}{\bibinfo{person}{S. {Rendle}}.}
  \bibinfo{year}{2010}\natexlab{}.
\newblock \showarticletitle{Factorization Machines}. In
  \bibinfo{booktitle}{\emph{2010 IEEE International Conference on Data
  Mining}}. \bibinfo{pages}{995--1000}.
\newblock


\bibitem[\protect\citeauthoryear{Rendle, Freudenthaler, Gantner, and
  Schmidt-Thieme}{Rendle et~al\mbox{.}}{2009}]%
        {BPR}
\bibfield{author}{\bibinfo{person}{Steffen Rendle}, \bibinfo{person}{Christoph
  Freudenthaler}, \bibinfo{person}{Zeno Gantner}, {and} \bibinfo{person}{Lars
  Schmidt-Thieme}.} \bibinfo{year}{2009}\natexlab{}.
\newblock \showarticletitle{BPR: Bayesian Personalized Ranking from Implicit
  Feedback}. In \bibinfo{booktitle}{\emph{Proceedings of the Twenty-Fifth
  Conference on Uncertainty in Artificial Intelligence}} (Montreal, Quebec,
  Canada) \emph{(\bibinfo{series}{UAI '09})}. \bibinfo{publisher}{AUAI Press},
  \bibinfo{address}{Arlington, Virginia, USA}, \bibinfo{pages}{452–461}.
\newblock
\showISBNx{9780974903958}


\bibitem[\protect\citeauthoryear{Rendle, Gantner, Freudenthaler, and
  Schmidt-Thieme}{Rendle et~al\mbox{.}}{2011}]%
        {MF}
\bibfield{author}{\bibinfo{person}{Steffen Rendle}, \bibinfo{person}{Zeno
  Gantner}, \bibinfo{person}{Christoph Freudenthaler}, {and}
  \bibinfo{person}{Lars Schmidt-Thieme}.} \bibinfo{year}{2011}\natexlab{}.
\newblock \showarticletitle{Fast Context-Aware Recommendations with
  Factorization Machines}. In \bibinfo{booktitle}{\emph{Proceedings of the 34th
  International ACM SIGIR Conference on Research and Development in Information
  Retrieval}} (Beijing, China) \emph{(\bibinfo{series}{SIGIR '11})}.
  \bibinfo{publisher}{Association for Computing Machinery},
  \bibinfo{address}{New York, NY, USA}, \bibinfo{pages}{635–644}.
\newblock
\showISBNx{9781450307574}


\bibitem[\protect\citeauthoryear{Sergeev and Balso}{Sergeev and Balso}{2018}]%
        {horovod}
\bibfield{author}{\bibinfo{person}{Alexander Sergeev} {and}
  \bibinfo{person}{Mike~Del Balso}.} \bibinfo{year}{2018}\natexlab{}.
\newblock \showarticletitle{Horovod: fast and easy distributed deep learning in
  TensorFlow}.
\newblock \bibinfo{journal}{\emph{CoRR}}  \bibinfo{volume}{abs/1802.05799}
  (\bibinfo{year}{2018}).
\newblock


\bibitem[\protect\citeauthoryear{{Shi}, {Hu}, {Zhao}, and {Yu}}{{Shi}
  et~al\mbox{.}}{2019}]%
        {HINsForRecommendation}
\bibfield{author}{\bibinfo{person}{C. {Shi}}, \bibinfo{person}{B. {Hu}},
  \bibinfo{person}{W.~X. {Zhao}}, {and} \bibinfo{person}{P.~S. {Yu}}.}
  \bibinfo{year}{2019}\natexlab{}.
\newblock \showarticletitle{Heterogeneous Information Network Embedding for
  Recommendation}.
\newblock \bibinfo{journal}{\emph{IEEE Transactions on Knowledge and Data
  Engineering}} \bibinfo{volume}{31}, \bibinfo{number}{2}
  (\bibinfo{year}{2019}), \bibinfo{pages}{357--370}.
\newblock


\bibitem[\protect\citeauthoryear{Su and Khoshgoftaar}{Su and
  Khoshgoftaar}{2009}]%
        {CF_Survey2}
\bibfield{author}{\bibinfo{person}{Xiaoyuan Su} {and} \bibinfo{person}{Taghi~M.
  Khoshgoftaar}.} \bibinfo{year}{2009}\natexlab{}.
\newblock \showarticletitle{A Survey of Collaborative Filtering Techniques}.
\newblock \bibinfo{journal}{\emph{Adv. in Artif. Intell.}}
  \bibinfo{volume}{2009}, Article \bibinfo{articleno}{4} (\bibinfo{date}{Jan.}
  \bibinfo{year}{2009}), \bibinfo{numpages}{1}~pages.
\newblock
\showISSN{1687-7470}


\bibitem[\protect\citeauthoryear{Tak\'{a}cs and Tikk}{Tak\'{a}cs and
  Tikk}{2012}]%
        {ALS_PR}
\bibfield{author}{\bibinfo{person}{G\'{a}bor Tak\'{a}cs} {and}
  \bibinfo{person}{Domonkos Tikk}.} \bibinfo{year}{2012}\natexlab{}.
\newblock \showarticletitle{Alternating Least Squares for Personalized
  Ranking}. In \bibinfo{booktitle}{\emph{Proceedings of the Sixth ACM
  Conference on Recommender Systems}} (Dublin, Ireland)
  \emph{(\bibinfo{series}{RecSys '12})}. \bibinfo{publisher}{Association for
  Computing Machinery}, \bibinfo{address}{New York, NY, USA},
  \bibinfo{pages}{83–90}.
\newblock
\showISBNx{9781450312707}


\bibitem[\protect\citeauthoryear{Tian, Zhang, Rang, Yang, and Zhan}{Tian
  et~al\mbox{.}}{2020}]%
        {RGCN}
\bibfield{author}{\bibinfo{person}{Anqi Tian}, \bibinfo{person}{Chunhong
  Zhang}, \bibinfo{person}{Miao Rang}, \bibinfo{person}{Xueying Yang}, {and}
  \bibinfo{person}{Zhiqiang Zhan}.} \bibinfo{year}{2020}\natexlab{}.
\newblock \showarticletitle{RA-GCN: Relational Aggregation Graph Convolutional
  Network for Knowledge Graph Completion}. In
  \bibinfo{booktitle}{\emph{Proceedings of the 2020 12th International
  Conference on Machine Learning and Computing}} (Shenzhen, China)
  \emph{(\bibinfo{series}{ICMLC 2020})}. \bibinfo{publisher}{Association for
  Computing Machinery}, \bibinfo{address}{New York, NY, USA},
  \bibinfo{pages}{580–586}.
\newblock
\showISBNx{9781450376426}


\bibitem[\protect\citeauthoryear{van~den Berg, Kipf, and Welling}{van~den Berg
  et~al\mbox{.}}{2017}]%
        {GCMC}
\bibfield{author}{\bibinfo{person}{Rianne van~den Berg},
  \bibinfo{person}{Thomas~N. Kipf}, {and} \bibinfo{person}{Max Welling}.}
  \bibinfo{year}{2017}\natexlab{}.
\newblock \showarticletitle{Graph Convolutional Matrix Completion}.
\newblock \bibinfo{journal}{\emph{CoRR}}  \bibinfo{volume}{abs/1706.02263}
  (\bibinfo{year}{2017}).
\newblock
\showeprint[arxiv]{1706.02263}


\bibitem[\protect\citeauthoryear{Wang, Wang, and Yeung}{Wang
  et~al\mbox{.}}{2015}]%
        {CollaborativeDeepLearning}
\bibfield{author}{\bibinfo{person}{Hao Wang}, \bibinfo{person}{Naiyan Wang},
  {and} \bibinfo{person}{Dit-Yan Yeung}.} \bibinfo{year}{2015}\natexlab{}.
\newblock \showarticletitle{Collaborative Deep Learning for Recommender
  Systems}. In \bibinfo{booktitle}{\emph{Proceedings of the 21th ACM SIGKDD
  International Conference on Knowledge Discovery and Data Mining}} (Sydney,
  NSW, Australia) \emph{(\bibinfo{series}{KDD '15})}.
  \bibinfo{publisher}{Association for Computing Machinery},
  \bibinfo{address}{New York, NY, USA}, \bibinfo{pages}{1235–1244}.
\newblock


\bibitem[\protect\citeauthoryear{Wang, He, Wang, Feng, and Chua}{Wang
  et~al\mbox{.}}{2019}]%
        {NeuralGraphCollaborativeFiltering}
\bibfield{author}{\bibinfo{person}{Xiang Wang}, \bibinfo{person}{Xiangnan He},
  \bibinfo{person}{Meng Wang}, \bibinfo{person}{Fuli Feng}, {and}
  \bibinfo{person}{Tat-Seng Chua}.} \bibinfo{year}{2019}\natexlab{}.
\newblock \showarticletitle{Neural Graph Collaborative Filtering}. In
  \bibinfo{booktitle}{\emph{Proceedings of the 42nd International ACM SIGIR
  Conference on Research and Development in Information Retrieval}} (Paris,
  France) \emph{(\bibinfo{series}{SIGIR'19})}. \bibinfo{publisher}{Association
  for Computing Machinery}, \bibinfo{address}{New York, NY, USA},
  \bibinfo{pages}{165–174}.
\newblock
\showISBNx{9781450361729}


\bibitem[\protect\citeauthoryear{Wu, DuBois, Zheng, and Ester}{Wu
  et~al\mbox{.}}{2016}]%
        {collabdenoise}
\bibfield{author}{\bibinfo{person}{Yao Wu}, \bibinfo{person}{Christopher
  DuBois}, \bibinfo{person}{Alice~X. Zheng}, {and} \bibinfo{person}{Martin
  Ester}.} \bibinfo{year}{2016}\natexlab{}.
\newblock \showarticletitle{Collaborative Denoising Auto-Encoders for Top-N
  Recommender Systems}. In \bibinfo{booktitle}{\emph{Proceedings of the Ninth
  ACM International Conference on Web Search and Data Mining}}.
\newblock


\bibitem[\protect\citeauthoryear{Wu, Pan, Chen, Long, Zhang, and Yu}{Wu
  et~al\mbox{.}}{2019}]%
        {survey_wu}
\bibfield{author}{\bibinfo{person}{Zonghan Wu}, \bibinfo{person}{Shirui Pan},
  \bibinfo{person}{Fengwen Chen}, \bibinfo{person}{Guodong Long},
  \bibinfo{person}{Chengqi Zhang}, {and} \bibinfo{person}{Philip~S. Yu}.}
  \bibinfo{year}{2019}\natexlab{}.
\newblock \showarticletitle{A Comprehensive Survey on Graph Neural Networks}.
\newblock \bibinfo{journal}{\emph{ArXiv}}  \bibinfo{volume}{abs/1901.00596}
  (\bibinfo{year}{2019}).
\newblock


\bibitem[\protect\citeauthoryear{Xu, Lian, Han, Li, Xu, and Xie}{Xu
  et~al\mbox{.}}{2019}]%
        {RecoGCN}
\bibfield{author}{\bibinfo{person}{Fengli Xu}, \bibinfo{person}{Jianxun Lian},
  \bibinfo{person}{Zhenyu Han}, \bibinfo{person}{Yong Li},
  \bibinfo{person}{Yujian Xu}, {and} \bibinfo{person}{Xing Xie}.}
  \bibinfo{year}{2019}\natexlab{}.
\newblock \showarticletitle{Relation-Aware Graph Convolutional Networks for
  Agent-Initiated Social E-Commerce Recommendation}. In
  \bibinfo{booktitle}{\emph{Proceedings of the 28th ACM CIKM}}.
\newblock


\bibitem[\protect\citeauthoryear{Yang, Ding, Zhou, Yang, Zhou, and Tang}{Yang
  et~al\mbox{.}}{2020}]%
        {NegSamp}
\bibfield{author}{\bibinfo{person}{Zhen Yang}, \bibinfo{person}{Ming Ding},
  \bibinfo{person}{Chang Zhou}, \bibinfo{person}{Hongxia Yang},
  \bibinfo{person}{Jingren Zhou}, {and} \bibinfo{person}{Jie Tang}.}
  \bibinfo{year}{2020}\natexlab{}.
\newblock \showarticletitle{Understanding Negative Sampling in Graph
  Representation Learning}. \bibinfo{pages}{1666--1676}.
\newblock


\bibitem[\protect\citeauthoryear{Ying, He, Chen, Eksombatchai, Hamilton, and
  Leskovec}{Ying et~al\mbox{.}}{2018}]%
        {PinSage}
\bibfield{author}{\bibinfo{person}{Rex Ying}, \bibinfo{person}{Ruining He},
  \bibinfo{person}{Kaifeng Chen}, \bibinfo{person}{Pong Eksombatchai},
  \bibinfo{person}{William~L. Hamilton}, {and} \bibinfo{person}{Jure
  Leskovec}.} \bibinfo{year}{2018}\natexlab{}.
\newblock \showarticletitle{Graph Convolutional Neural Networks for Web-Scale
  Recommender Systems}. In \bibinfo{booktitle}{\emph{KDD}}.
\newblock


\bibitem[\protect\citeauthoryear{Zhang, Yuan, Lian, Xie, and Ma}{Zhang
  et~al\mbox{.}}{2016}]%
        {CollborativeKnowledgeBase}
\bibfield{author}{\bibinfo{person}{Fuzheng Zhang},
  \bibinfo{person}{Nicholas~Jing Yuan}, \bibinfo{person}{Defu Lian},
  \bibinfo{person}{Xing Xie}, {and} \bibinfo{person}{Wei-Ying Ma}.}
  \bibinfo{year}{2016}\natexlab{}.
\newblock \showarticletitle{Collaborative Knowledge Base Embedding for
  Recommender Systems}. In \bibinfo{booktitle}{\emph{Proceedings of the 22nd
  ACM SIGKDD International Conference on Knowledge Discovery and Data Mining}}
  (San Francisco, California, USA) \emph{(\bibinfo{series}{KDD '16})}.
  \bibinfo{publisher}{Association for Computing Machinery},
  \bibinfo{address}{New York, NY, USA}, \bibinfo{pages}{353–362}.
\newblock
\showISBNx{9781450342322}


\bibitem[\protect\citeauthoryear{Zhang and Chen}{Zhang and Chen}{2020}]%
        {IGMC}
\bibfield{author}{\bibinfo{person}{Muhan Zhang} {and} \bibinfo{person}{Yixin
  Chen}.} \bibinfo{year}{2020}\natexlab{}.
\newblock \showarticletitle{Inductive Matrix Completion Based on Graph Neural
  Networks}. In \bibinfo{booktitle}{\emph{ICLR}}.
\newblock


\bibitem[\protect\citeauthoryear{Zhang, Yao, Sun, and Tay}{Zhang
  et~al\mbox{.}}{2019}]%
        {DL_based_recommender_system}
\bibfield{author}{\bibinfo{person}{Shuai Zhang}, \bibinfo{person}{Lina Yao},
  \bibinfo{person}{Aixin Sun}, {and} \bibinfo{person}{Yi Tay}.}
  \bibinfo{year}{2019}\natexlab{}.
\newblock \showarticletitle{Deep Learning Based Recommender System: A Survey
  and New Perspectives}.
\newblock \bibinfo{journal}{\emph{ACM Comput. Surv.}} \bibinfo{volume}{52},
  \bibinfo{number}{1}, Article \bibinfo{articleno}{5} (\bibinfo{date}{Feb.}
  \bibinfo{year}{2019}), \bibinfo{numpages}{38}~pages.
\newblock
\showISSN{0360-0300}


\bibitem[\protect\citeauthoryear{Zheng, Liu, Shi, Zhuang, Li, and Wu}{Zheng
  et~al\mbox{.}}{2017}]%
        {douban}
\bibfield{author}{\bibinfo{person}{Jing Zheng}, \bibinfo{person}{Jian Liu},
  \bibinfo{person}{Chuan Shi}, \bibinfo{person}{Fuzhen Zhuang},
  \bibinfo{person}{Jingzhi Li}, {and} \bibinfo{person}{Bin Wu}.}
  \bibinfo{year}{2017}\natexlab{}.
\newblock \showarticletitle{Recommendation in heterogeneous information network
  via dual similarity regularization}.
\newblock \bibinfo{journal}{\emph{International Journal of Data Science and
  Analytics}}  \bibinfo{volume}{3} (\bibinfo{date}{02} \bibinfo{year}{2017}).
\newblock


\end{thebibliography}

%%
%% If your work has an appendix, this is the place to put it.
%\appendix

\end{document}